\documentclass[twocolumn]{aa} 

\usepackage[varg]{txfonts}
\usepackage{graphicx}
\usepackage{txfonts}
\usepackage{multirow}
\usepackage{xcolor}
\usepackage{tabularx}
\usepackage{array}
\usepackage{rotating}
\usepackage{makecell}
\usepackage{tablefootnote}
\usepackage{booktabs}
\usepackage{threeparttable}
\usepackage{siunitx}
\usepackage{soul}
\usepackage[utf8]{inputenc}
\usepackage{hyperref}
\def\ew{$W_{2796}$}

\def\usmgii{$W_{2796} \ge 3\AA$}
\def\ewfeii{$W_{2600}$}

\def\kms{km~s$^{-1}$}

\def\mgii{Mg~{\sc ii}~} 
\def\oii{O~{[\sc ii}]~} 
 
\def\mgiia{Mg~{\sc ii} $\lambda$2796} 
 
\def\mgiiab{Mg~{\sc ii} $\lambda\lambda$2796,2803~} 

\newcommand{\romannumber}[1]{\MakeUppercase{\romannumeral #1}}

\def\feii{Fe~{\sc ii}~} 
 
\def\feiia{Fe~{\sc ii} $\lambda$2600~} 

\def\oiii{[O~{\sc iii}]} 
\def\oiiab{[O~{\sc ii}] $\lambda\lambda$3727,3729}  
  
\def\oii{[O~{\sc ii}]}

\def\o3o2{[O~{\sc iii}]/[O~{\sc ii}]}

\DeclareUnicodeCharacter{02BC}{'}
\def\numqsosightlinesdrsixteen{107,263}
\def\numabsorbersdssdrsixteen{159,524}
\def\numuniqueabsorbersdssdrsixteen{130,596}
\def\numuniqueusmgiiabsorbersdssdrsixteen{4,381}

\def\numqsosightlinesusmgiihsc{630}
\def\numabsorberusmgiihsc{636}
\def\numgoodabsorberusmgiihsc{583}
\def\usmgiipopulationpercentage{3}
\def\kpconepfivearcsecavgz{12}
\def\kpconepfivearcseczeropfivez{9}
\def\kpconepfivearcsectwopfivez{13}
\def\bosscoveragepercentage{89}
\def\bosscoveragegalonearcsecpercentage{77}
\def\oiidetectiontwosigmapercent{29}
\def\oiidetectiononepfivesigmapercent{44}
\def\hintproximdetectionpercent{9.7}
\def\darkexampleoiisigma{3.2}
\def\totalnumberofusmgiiabsorbers{418}
\def\totalnumberofusmgiisightlines{412}
\def\usmgiisightlinesclosebydetection{168}
\def\usmgiiabsorbersclosebydetection{170}
\def\usmgiiclearsightlines{175}
\def\usmgiiclearsightlinesabsorbers{178}

\def\totaldiscardsightlines{69}
\def\closebydetectionthreesigmaminfourfilter{156}
\def\closebydetectionoiicovered{170}
\def\closebydetectionoiidetected{50}
\def\closebydetectionoiidetectedthreesigma{27}
\def\closebydetectionoiinotdetected{120}
\def\closebydetectionvideosightlines{3}
\def\closebydetectionvikingsightlines{29}
\def\closebydetectionvhssightlines{58}
\def\closebydetectiondecamwisesightlines{75}
\def\detection{136}
\def\detectionpercent{40}
\def\detectionpercentminfourfilter{38}
\def\nondetection{204}
\def\nondetectionpercent{60}
\def\totalminimumfourfilteroroiidetected{340}
\def\nondetectionpercentminfourfilter{62}
\def\categoryonedetection{130}
\def\detectionoiidetected{50}

\def\detectionoiinotdetectednotcovered{86}
\def\detectionoiicovered{130}
\def\detectionoiinondetectednotcovered{86}

\def\detectionphotozmatched{82}
\def\detectionzfixgoodfitadditional{48}
\def\allabszmin{0.4}
\def\allabszmax{1.7}

\def\alldetectionseparcsecmax{2.9}
\def\detectiongoodpa{40}
\def\detectiongoodpaewfourangstrom{3}
\def\ipstartkpc{5.2}
\def\ipendkpc{23}
\def\ipmediankpc{11.4}
\def\totaldiskwindsource{38}
\def\totalwindsource{26}
\def\totaldisksource{12}
\def\megaflowabsorbercount{8}

\def\windsubsetmedianew{3.59}
\def\windsubsetstdew{0.08}
\def\disksubsetmedianew{3.28}
\def\disksubsetstdew{0.06}
\def\diskwindewpvalue{0.084}

\def\detectiongoodpamajorpercent{30}
\def\detectiongoodpamajor{12}
\def\detectiongoodpaminorpercent{52}
\def\detectiongoodpaminor{21}
\def\detectiongoodpainbetweenmajorminorpercent{18}
\def\detectiongoodpainbetweenmajorminor{7}
\def\averagessfrdiskwind{-8.80}
\def\averagefeiibymgiidiskwind{0.62}
\def\averagefeiibymgiidetection{0.64}
\def\stdfeiibymgiidetection{0.01}
\def\averagefeiibymgiinondetection{0.57}
\def\stdfeiibymgiinondetection{0.01}
\def\kstestfeiibymgiipvalue{0.004}
\def\ewdoverallfitbestfitalpha{0.628}
\def\ewdoverallfitbestfitalphaupper{0.021}
\def\ewdoverallfitbestfitalphalower{0.019}
\def\ewdoverallfitbestfitbeta{-0.018}
\def\ewdoverallfitbestfitbetaupper{0.001}
\def\ewdoverallfitbestfitbetalower{0.001}
\def\minsfrdetected{0.01}
\def\meansfrdetected{30.72}
\def\maxsfrdetected{340}

\def\tauewvssfrstat{0.10}
\def\tauewvssfrpvalue{0.10}
\def\tauewvsmstarstat{0.13}
\def\tauewvsmstarpvalue{0.03}
\def\tauewfeiivssfrstat{0.12}
\def\tauewfeiivssfrpvalue{0.04}

\def\tauewfeiimgiivssfrstat{0.11}
\def\tauewfeiimgiivssfrpvalue{0.05}

\def\minstellarmassdetected{8.65}
\def\maxstellarmassdetected{11.67}
\def\averagestellarmass{9.90}
\def\mediansfrbelowaveragestellarmass{5.48}
\def\stdsfrbelowaveragestellarmassdex{0.7}
\def\mediansfraboveaveragestellarmass{30.29}
\def\stdsfraboveaveragestellarmassdex{1.6}

\def\starburstpercent{21}
\def\excessgalaxiesperkpcsquare{0.00031}

\def\singlehostabsorberincourspercent{14}
\def\doublehostabsorberincourpercent{17}
\def\multihostabsorberincourspercent{66}
\def\totalcategoryaoiisnr{13.7}

\def\totalcategoryboiisnr{4.2}
\def\categoryoneboiisnr{3.1}
\def\totalcategoryaboiisnr{11.3}
\def\categoryoneaboiisnr{10.5}
\def\totalclearoiisnr{2.8}
\def\numbercone{170}
\def\numberconeone{156}
\def\numberconeoneone{82}
\def\numberconeonetwo{48}
\def\numberconeonethree{26}
\def\numberconetwo{14}
\def\numberctwo{248}
\def\numberctwoone{178}
\def\numberctwotwo{70}
\def\categoryonea{44}
\def\categoryoneb{86}

\def\categorytwo{26}
\def\categorythreea{6}
\def\categorythreeb{8}

\def\categoryclear{178}
\def\categorydiscard{70}

\begin{document}

   \title{Baryonic Ecosystem in Galaxies (BEINGMgII). Host Galaxies of Ultra-strong \mgii Absorbers in Subaru Hyper Suprime-Cam Survey}


   \author{Ravi Joshi
          \inst{1}, 
          Sarbeswar Das\inst{1}, 
          Michele Fumagalli\inst{2,3},
          Matteo Fossati\inst{2,4}, 
          C\'eline P\'eroux\inst{5,6},
          Reena Chaudhary\inst{1}, 
          Hassen M. Yesuf\inst{7,8,9}, 
          \and
          Luis C. Ho\inst{9,10}
          }

   \institute{Indian Institute of Astrophysics (IIA), Koramangala, Bangalore 560034, India \\     \email{rvjoshirv@gmail.com}
   \and
             Universit\'a degli Studi di Milano-Bicocca, Dip. di Fisica G. Occhialini, Piazza della Scienza 3, 20126 Milano, Italy
             \and
             INAF - Osservatorio Astronomico di Trieste, via G.B. Tiepolo 11, I-34143 Trieste, Italy
             \and 
             INAF - Osservatorio Astronomico di Brera, Via Brera 28, 21021, Milano, Italy
             \and
             European Southern Observatory, Karl-Schwarzschildstrasse 2, D-85748 Garching bei Munchen, Germany
             \and 
             Aix Marseille Universit\'e, CNRS, LAM (Laboratoire d’Astrophysique de Marseille) UMR 7326, F-13388 Marseille, France
             \and 
             Key Laboratory for Research in Galaxies and Cosmology, Shanghai Astronomical Observatory, Chinese Academy of Sciences, 80 Nandan Road, Shanghai 200030, People’s Republic of China
            \and 
            Kavli Institute for the Physics and Mathematics of the Universe (WPI), The University of Tokyo Institutes for Advanced Study (UTIAS), The University of Tokyo, 5-1-5 Kashiwanoha, Kashiwa-shi, Chiba 277-8583, Japan
             \and 
             Kavli Institute for Astronomy and Astrophysics, Peking University, Beijing 100871, Peopleʼs Republic of China
             \and
             Department of Astronomy, School of Physics, Peking University, Beijing 100871, Peopleʼs Republic of China
             }
   \date{Received July 26, 2024; accepted July 26, 2024}

 \titlerunning{BEINGMgII- Host galaxies of Ultra-Strong \mgii Absorbers}
  \authorrunning{Joshi, et al.}
  \abstract
   {We study the galaxies hosting ultra-strong \mgii (USMgII) absorbers at small impact parameters of $\sim$\SI{2}{\arcsecond} (5 - 20 kpc), spanning a redshift range of $\allabszmin\ \le z \le \allabszmax$, using deep, high-resolution images from Hyper Suprime-Cam Subaru Strategic Survey and spectra from SDSS survey.}
   {To explore the physical origin of USMgII absorbers and characterize the associated galaxies.}
   {We performed a galaxy spectral energy distribution fitting using optical and near-IR multi-band data to identify the potential absorber host galaxies. Further, we search for \oii\ nebular emission line from absorber galaxy in the SDSS fiber spectra.}
   {From a total of \totalnumberofusmgiiabsorbers\ USMgII absorbers with \ew $\ge 3 $\AA\, along \totalnumberofusmgiisightlines\ quasar sightlines, we detect \detectionoiidetected\ galaxies based on \oiiab\ nebular emission detected at $\ge 2\sigma$ level. Utilizing the \oii\ emission from the stacked spectrum and employing the best-fit galaxy SED template, we further identify \detectionoiinotdetectednotcovered\ galaxies, leading to a total of \detection\ bona fide USMgII galaxies. With a prerequisite of having a minimum of four HSC passbands available, we find a detection rate of $\sim$\detectionpercentminfourfilter\% at an average impact parameter of \ipmediankpc\ kpc. We find that galaxies hosting USMgII systems are typically star-forming main sequence galaxies, with \starburstpercent\% exhibiting a starburst nature. The non-zero \oii\ emission along the `clear' sightlines, with no stellar counterpart, hints that the USMgII absorbers may likely emanate from the unseen faint galaxies near the quasar. The USMgII absorbers preferentially align along the major and minor axes of the galaxy, which suggests that they originate in the disk or large-scale wind. We show that the distribution of \ew\ as a function of impact parameter indicates a discernible radial dependence for the `disk' and `wind' subsets, with the observed large scatter in \ew\ potentially attributed to large-scale outflows. The quasar sightline hosting USMgII systems show a factor three higher galaxy surface density at impact parameters of $\lesssim 50$kpc, highlights the multiple pathways giving rise to USMgII absorption.}
   {}

  \maketitle
%

\section{Introduction}
\label{sec:intro}
Gas accretion and galactic winds are firmly established as essential drivers for the growth of galaxies. Simulations suggest that the galactic winds regulate star formation by expelling gas and heating the interstellar and circumgalactic medium (CGM), thereby influencing the rate and efficiency of star formation within galaxies \citep{Carr2023ApJ...949...21C}. Besides shaping the galaxy luminosity function, they also significantly influence the stellar mass-metallicity relation \citep{Tremonti2004ApJ...613..898T} and inject a large amount of dust and metals into the galactic halo and intergalactic medium (IGM) \citep{Oppenheimer2006MNRAS.373.1265O,Peroux2020MNRAS.499.2462P}. \par
The widespread presence of outflows is evidenced by a commonly observed blue-shifted resonance-line absorption in the spectra of normal, star-forming galaxies with high specific star formation rates (sSFR) \citep{Rupke2019Natur.574..643R,Kehoe2024arXiv240607621K}. In starburst galaxies, such winds are seen as strong \mgii absorption with velocities $\gtrsim 1000$~\kms, attributed to the superwinds \citep{Heckman2000ApJS..129..493H}. Such velocities are comparable to or exceed the escape velocity from L$^{\star}$ galaxies. However, wind properties such as mass outflow rate and mass loading factor are poorly constrained due to the large uncertainty associated with the distance of absorbing gas, metallicity, and the ionization parameter \citep{Veilleux2005ARA&A..43..769V}.\par

Quasar absorption-line studies, tracing low-ion absorption in the spectrum of a bright background source, provide information about the gas flows in and around galaxies in a luminosity-unbiased manner \citep{Steidel2010ApJ...717..289S}.  
The majority of quasar-galaxy pair research has focused on the resonant \mgiiab\ doublet lines, which are produced from a photoionized gas at a temperature of $10^4 K$, and are extensively accessible over $0.3 \le z \le 2.3$ from the ground \citep{Tumlinson2017ARA&A..55..389T}. This transition is extremely sensitive to the presence of the interstellar medium and the CGM 
\citep{Tumlinson2017ARA&A..55..389T, Feltre2018A&A...617A..62F}.  Such investigations, particularly of strong \mgii absorbers with \ew $> 0.7$\AA, have revealed that the galaxy halo of most $sub-L^{\star}$ galaxies contains metal-enriched gas that extends to projected distances of $> 100$ kpc \citep{Chen2010ApJ...714.1521C,Nielsen2013ApJ...776..115N}. The \mgii gas is found to have patchy gaseous cross-sections with a high covering fraction of $\sim 70\%$ out to $\sim 50$kpc for \ew\ $> 0.3$ \AA\ \citep{Dutta2020MNRAS.499.5022D}. The absorption strength strongly depends on the stellar mass and the distance from the galaxy.  A bi-modality is observed in the distribution of  both neutral hydrogen and metal-enriched gas as well as its metallicity as a function of azimuthal angle, intriguingly also seen in emission (\citealt{Guo2023Natur.624...53G}, but see, \citealt{Dutta2023MNRAS.522..535D} for isotropic signal), which supports their wind origin \citep{Kacprzak2014ApJ...792L..12K,Lan2018ApJ...866...36L,Zabl2021MNRAS.507.4294Z,Weng2023MNRAS.523..676W}. \par

A rare class of \mgii absorbers are ultra-strong \mgii systems defined as \ew $> 3$\AA\ and comprise only \usmgiipopulationpercentage\% of MgII absorber population. Unlike the kinematic spread of $\sim100$~\kms\ seen in Galactic \mgii absorbers in quasar spectra, the USMgII absorbers exhibit strong absorption over a velocity spread of several hundred \kms, which is likely to originate from the galactic superwind \citep{Bond2001ApJ...562..641B} or gas dynamics of the intra-group medium \citep{Gauthier2013MNRAS.432.1444G}. The direct evidence of large-scale outflows in USMgII systems is seen in high-redshift starburst galaxies \citep{Nestor2011MNRAS.412.1559N, Rubin2010ApJ...712..574R}. \citet{Menard2011MNRAS.417..801M}  found that the strong \mgii trace a substantial fraction of the global \oii\ luminosity density, and most likely trace the galaxies with vigorous star formation \citep[see also,][] {Joshi2017MNRAS.471.1910J,Joshi2018MNRAS.476..210J}.  The association of strong \mgii absorbers with the bluer galaxies close to quasar line-of-sight is also revealed in coadded images from the SDSS survey \citep{Zibetti2007ApJ...658..161Z}. However, the true nature of USMgII absorber host galaxies and their origin remains unclear. \par

In recent efforts to study the absorber galaxy connection using advanced integral field spectrographs, mapping the galaxies over large impact parameters ($\sim$ 200 kpc), the majority of the \mgii absorbers are commonly observed to be associated with multiple galaxies. The large scatter observed in absorbing galaxy properties might indicate that the observed absorption essentially may not only map the outflows but may also originate from the multiple halos of a galaxy group \citep{Bielby2017MNRAS.468.1373B, Peroux2017MNRAS.464.2053P,Fossati2019MNRAS.484.2212F,Dutta2020MNRAS.499.5022D}, intra-group medium \citep{Gauthier2013MNRAS.432.1444G} and/or cool stripped gas from environmental processes \cite{Dutta2020MNRAS.499.5022D}. The notion that the absorption is caused by galaxies at the close impact parameter, however, remains viable because it is highly challenging to find such faint galaxies in the glare of a bright background quasar. Most recently, \citet{Guha2024MNRAS.527.5075G} have detected 28 USMgII absorber host galaxies at an average impact parameter of 19 kpc, over  $0.6 \le z \le 0.8$.  They show that $\sim$29\% of USMgII absorbers with a wide velocity spread of more than 300\kms\ originate from gas flows (infall/outflow) in isolated massive galaxies, while  $\ge 21$\% of these systems emanate from the galaxy interaction. Moreover, they found that USMgII host galaxies exhibit slightly lower ongoing star-formation rates compared to main sequence galaxies with the same stellar mass, suggesting a transition from star-forming to quiescent states  \citep[see,][]{Guha2022MNRAS.513.3836G}. Therefore, it is imperative to investigate the possible origin of USMgII systems, their connection to galaxies, and their overall contribution to the global SFR density.

This paper is organized as follows. Section 2 describes our sample and the analysis. In Section 3, we examine the absorber galaxy association and the nature of galaxies associated with UGMgII absorbers.  The discussion and conclusions of this study are summarized in Section 4. Throughout, we have assumed the flat Universe with $H_0$ = 70 $\rm km\ s^{-1}\ Mpc^{-1}$, $\Omega_m$ = 0.3 and $\Omega_\Lambda$= 0.7.

\section{Sample Selection}

To probe the baryonic ecosystem in galaxies producing strong \mgii absorption (BEINGMgII), we exploit the largest \mgii absorber catalog from \citet{Anand2022MNRAS.513.3210A}, consisting of \numqsosightlinesdrsixteen\ quasar sightlines hosting a total of \numabsorbersdssdrsixteen\ \mgii absorbers. We consider the absorbers within 500 \kms\ as a single system, which resulted in \numuniqueabsorbersdssdrsixteen\ absorbers, of which \numuniqueusmgiiabsorbersdssdrsixteen\ are USMgII absorbers. To search for the possible faint absorber host galaxies, we have utilized deep multi-band ($g, r, i, z, y$) imaging data from the Hyper Supreme-Cam Subaru Strategic Program (HSC-SSP) Survey \citep{Aihara2022PASJ...74..247A}. The HSC-SSP offers high-resolution imaging data (median seeing of $\sim$ \SI{0.6}{\arcsecond}, depth of $ r \sim 26 $ \rm mag at $5\sigma$) over a wide sky region of about 1400 sq. degree. Among these \numuniqueusmgiiabsorbersdssdrsixteen\ USMgII absorbers, we found \numqsosightlinesusmgiihsc\ distinct quasar sightlines (hosting \numabsorberusmgiihsc\ absorbers) covered in HSC-SSP footprints.

Taking advantage of the multi-epoch spectrum from the Sloan Digital Sky Survey (SDSS), we generated the higher signal-to-noise ratio coadded spectrum and remeasured the \ew.  The absorbers with \mgiiab\ doublet ratio outside the optically thin ($W_{2796}/W_{2803} \approx$ 2) and optically thick ($W_{2796}/W_{2803} \approx$ 1) limits were then flagged for visual inspection, resulted in \numgoodabsorberusmgiihsc\ absorbers with \ew $>$ 3\AA. The marginal difference in the number of USMgII systems is primarily due to the removal of poor SNR systems. Finally, to ensure the coverage of key spectral features (e.g., $\rm D_{4000}$) to identify the absorber host galaxy within HSC passbands, we only select the absorbers with redshift $\leq 1.7$, led to a final sample of \totalnumberofusmgiiabsorbers\ USMgII systems along \totalnumberofusmgiisightlines\ sightlines.  \par

\section{Analysis}
\subsection{Searching for the potential host in HSC images}
\label{lab:hscdetect}

\begin{figure*}
    \centering
    \includegraphics[width=0.95\textwidth]{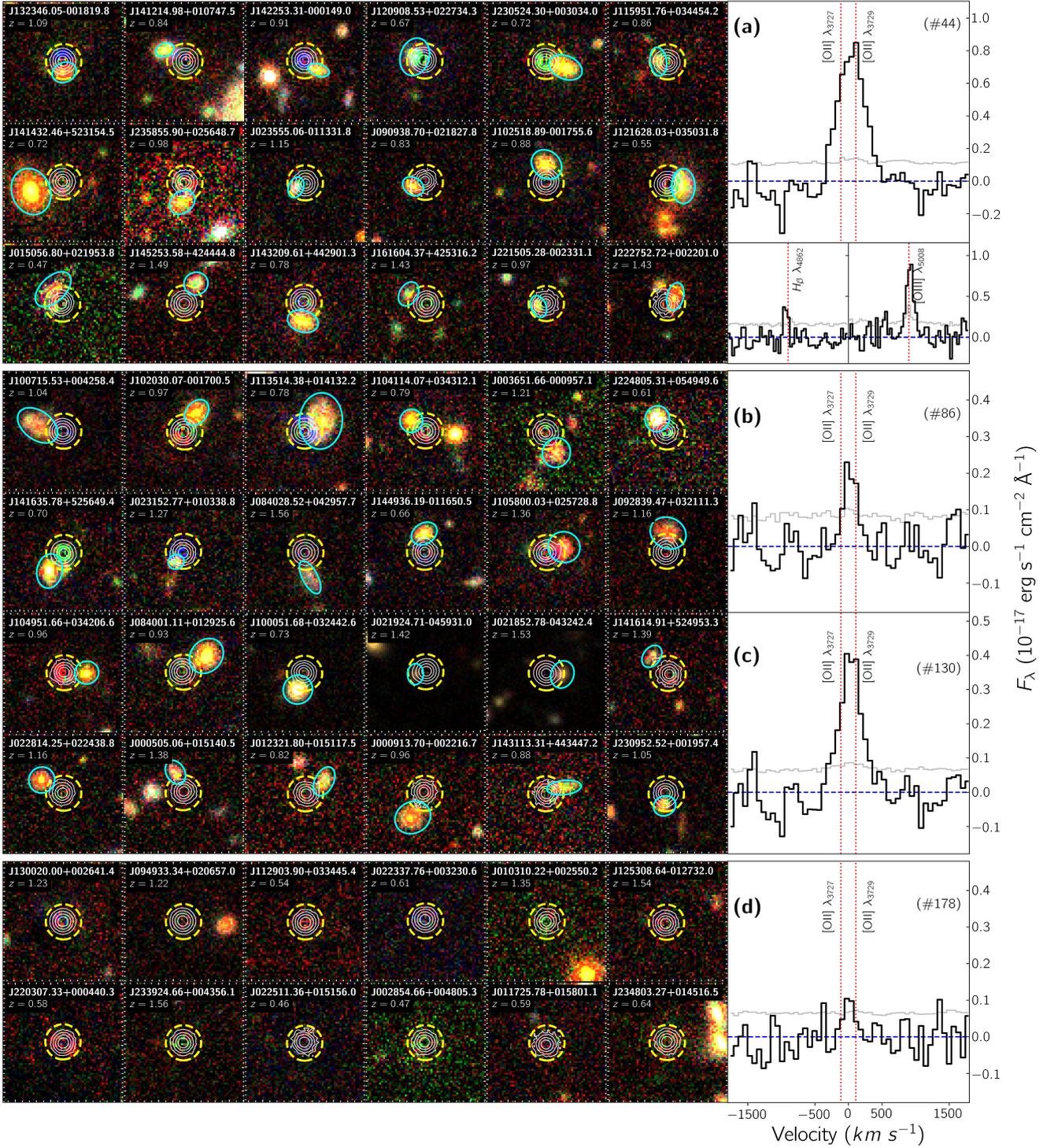}
    \caption{ The postage stamp HSC color composite images, centered on the quasar, are depicted with gray contours. The SDSS fiber, with a radius of \SI{1.5}{\arcsecond}, is indicated by dashed circles, while the cyan-colored aperture highlights the USMgII host galaxy. The first three rows exhibit an example set of potential host detections at close quasar proximity for USMgII systems with direct \oii\ detection (at $2\sigma$ level), followed by \oii\ detections $< 2\sigma$ level in rows 4-7. The bottom two rows display cases with no detection around the quasar. Additionally, the right panels showcase the stacked \oii\ emission line profiles and the grey curve shows the associated error.}
    \label{fig:collage}
\end{figure*}

At the survey depth of  $r \sim 26\ mag$ the quasar host galaxies are found to be prominent even at $z \sim 1$ \citep{Ishino2020PASJ...72...83I}. Therefore, we model the quasar with a point spread function (PSF) along with a S\'ersic profile representing the quasar host galaxy. Here, we utilized the differential evolution global optimization \citep{Storn1997} to achieve the best-fit model. Furthermore, we have implemented a regularization strategy to optimize the residuals. Most of the sample exhibits a well-modeled quasar profile, as depicted in Fig. \ref{fig:collage}, except for a few of the quasars with a saturated PSF or bad image quality.

Following the quasar PSF removal for all the filters, we obtained a deeper coadded image by median stacking all the bands using SWarp\footnote{\href{https://github.com/astromatic/swarp}{https://github.com/astromatic/swarp}} \citep{Bertin2002ASPC..281..228B}. Subsequently, we extract the sources from the coadded image frame using Source Extractor \citep{Bertin1996A&AS..117..393B}. Here, we demand at least five contiguous pixels be detected at 3 $\sigma$ levels. At first, we considered the apertures with consistent detection across at least three bands to avoid false-positive detection due to poor quasar removal. We have further performed a visual check of all the sightlines, in particular, to scrutinize the sources with limited (one or two) bands and images affected by saturation or improper PSF modeling.  \par 

Finally, for the absorber galaxy association, we restrict our search to objects for which twice the effective radius of the galaxy (comprising $\sim$75\% of the galaxy luminosity) lies inside the SDSS fiber radius of 1.5 arcsec (see below Sec~\ref{lab:oii}). It offers the advantage of identifying the absorber host galaxy based on strong nebular emission imprinted on the SDSS fiber spectra of the quasar. For a redshift range of $0.5 < z < 2$, the above search radius of 1.5 arcsec corresponds to a physical size of $\sim$\kpconepfivearcseczeropfivez\ --\kpconepfivearcsectwopfivez\ kpc. Over these redshift ranges, considering an average size evolution of $\sim 7 - 4$ kpc for a typical star-forming galaxy 
 with $\rm M_{\star} \sim 10^{10} M_{\odot}$ \citep{vanderwel2014ApJ...788...28V}, sets an upper limit on the impact parameters at $\sim17$kpc. Note that, based on the anti-correlation between \ew\ and $\rho$, a USMgII absorber host galaxy is expected to be observed at impact parameters of $\rm \lesssim 5 ~kpc$ \citep{Nielsen2013ApJ...776..115N}. In addition, the average impact parameter of USMgII systems in a handful of studies is found to be  $\rm \sim 20 ~kpc$ (see Section~\ref{sec:intro}) which indicates that our adopted search radius is adequate to trace the USMgII galaxies close to the quasar.

Among \totalnumberofusmgiisightlines\ sightlines, we detect a likely potential host galaxy in the deep HSC coadded images at $3\sigma$ level, along \usmgiisightlinesclosebydetection\ sightlines within the defined search radius. In the section that follows, we investigate their connection with the \usmgii\ systems. Additionally, we classify $\usmgiiclearsightlines$ sightlines,  hosting \categoryclear\ absorbers, with at least 4 HSC passbands and no stellar counterpart as `clear' sightlines. The remaining \totaldiscardsightlines\ sightlines without a stellar counterpart detected within the search radius and having the insufficient number of HSC passbands to be categorized as `clear' sightlines are excluded from further analysis. The summary of the detected galaxies, number of filters with source detection, sed-fitting results, and the respective categories are listed in Table~\ref{tab:usmgiisummary}. 

\subsection{[OII] nebular emission line within SDSS fiber}
\label{lab:oii}

The SDSS fiber-fed spectroscopic observations utilize a finite fiber of diameter \SI{3}{\arcsecond}  (SDSS-DR7; \citealt{Abazajian2009ApJS..182..543A}) and \SI{2}{\arcsecond} (SDSS-DR12: BOSS; \citealt{Alam2015ApJS..219...12A}) which register photons from all the objects that fall within the fiber along our line of sight \citep{Noterdaeme2010MNRAS.403..906N,Joshi2017MNRAS.471.1910J}.  To detect the \oiiab\ nebular emission, when available, we utilize the multi-epoch SDSS or BOSS spectrum and generate the  
 coadded higher SNR spectrum for the SDSS or BOSS spectrograph separately. For each USMgII absorber, we search for the \oii\ nebular emission lines at the expected location for $z_{abs}$ in
the continuum subtracted spectrum.  To get the best quasar continuum, we have used the iterative B-spline fitting along with the median smoothing function. Next, using FELIS\footnote{\href{https://github.com/kasperschmidt/FELIS}{https://github.com/kasperschmidt/FELIS}} \citep{Schmidt2021zndo...5131705S}, we cross-correlate a \oii\ nebular emission line template, generated from the SDSS galaxy template, at the absorption redshift of USMgII absorber, within $\pm 500 \rm km\ s^{-1}$. The FELIS uses
the model template (normalized to an integrated flux of 1) and apply a flux scaling to the template
to obtain the best match by providing the $S/N$ for the minimized $\chi^2$. 
Finally, we also performed a visual inspection to avoid any false positives due to poor sky subtraction.

For \usmgiisightlinesclosebydetection\  ($\usmgiiabsorbersclosebydetection$ absorbers) out of \totalnumberofusmgiisightlines\ sightlines, where we have detected a potential host galaxy in the deep HSC images, the \oii\ nebular emission line is covered within the SDSS spectral coverage of 3600-10,400\AA. We further note that \bosscoveragepercentage\% of quasar sightlines have BOSS spectra obtained with a fiber radius of 1 arcsec, however, for the majority (\bosscoveragegalonearcsecpercentage\%) of these systems, at least 2$R_e$ of the galaxies fall inside the fiber, increasing the likelihood of  \oii\ detection (see above).  We consider a \oii\ detection threshold of $2\sigma$, resulting in a total detection of \closebydetectionoiidetected\  USMgII absorber host galaxies (see Table~\ref{tab:usmgiisummary}). Among these systems, \oii\ nebular emission is detected at $\ge 3 \sigma$ level for \closebydetectionoiidetectedthreesigma\ USMgII systems. In Fig~\ref{fig:collage} (top three rows), we show examples where the galaxy is visible within the fiber radius of \SI{1.5}{\arcsecond}. A strong \oii\ emission is detected at $\sim \totalcategoryaoiisnr\ \sigma$ level in the stacked spectrum. For this subset, the other nebular emission, such as H$\beta$ and \oiii\ falls in the sky region, but we see these features in the coadded spectrum (see Fig.\ref{fig:collage}, last column).

Note that besides the strength of the \oii\ nebular emission line, its detection also depends on the quasar continuum flux level \citep[see,][]{Noterdaeme2010MNRAS.403..906N}. Thus, for \closebydetectionoiinotdetected\ out of \closebydetectionoiicovered\ systems, where the \oii\ SNR is at  $< 2\sigma$ level, we further investigate whether the sources detected in HSC images at quasar proximity are genuine USMgII host galaxies. For this, we generate the spectral stack to detect the \oii\ nebular emission and detect an emission line at $ \totalcategoryboiisnr \sigma$ level. We detect a strong \oii\ emission at $\totalcategoryaboiisnr\ \sigma$ level if we consider all the  \closebydetectionoiicovered\ systems, with and without individual \oii\ detection. This reiterates that most of our detections are genuine USMgII absorber host galaxies.

Next, ascertaining the availability of minimum 4 HSC passbands, we classified  \usmgiiclearsightlines\ sightlines, with no potential host galaxy detected within the search radius of \SI{1.5}{\arcsecond}, as a  `clear' sightlines. For this subset, we detect a non-zero flux at the \oii\ position, with a marginal detection at \totalclearoiisnr\ $\sigma$ level (see Fig~\ref{fig:collage}, panel d). It may likely be contributed by the extended \oii\ emission from a distant galaxy \citep{Perrotta2024arXiv240910013P}, a galaxy group \citep{Chen2019ApJ...878L..33C}, or suggests the presence of nearby faint galaxies producing the USMgII absorbers (see Section~\ref{lab:detectionfrac}). The clean quasar subtracted images are shown in the bottom two rows of Fig~\ref{fig:collage}. The summary of USMgII systems with \oii\ detection is listed in column 4 of Table~\ref{tab:usmgiisummary}.

\begin{table}[h!]
\caption{Source detection summary for \mgii\ absorbers based on photometry}
\label{tab:usmgiisummary}
\setlength{\tabcolsep}{0.15cm}
\centering
\scriptsize
\renewcommand{\arraystretch}{1.5}
\begin{tabular}{ 
        m{0.3cm} | m{1.5cm}  m{1.6cm}  m{1.9cm} | m{1.5cm}
    } 
    \toprule
     \multicolumn{1}{c}{} & Detection proximity (HSC) & No. of bands    \newline with $3\sigma$ Detection & SED Modelling \newline Results & \oii\ SNR \newline based Category \\ 
    \midrule
    \midrule
        \multirow{5}{*}{\centering\rotatebox[origin=c]{90}{ALL ABSORBERS {\tiny(\totalnumberofusmgiiabsorbers)}}} & 
        \multirow{4}{1.5cm}{\centering $2R_e \le \SI{1.5}{\arcsecond}$ {\tiny(\numbercone)}} & 
        \multirow{3}{1.2cm}{\centering $\ge 4$ {\tiny(\numberconeone)}} & {\centering $z_{phot}~\sim z_{abs}$ {\tiny(\numberconeoneone)}} &  
        \multirow{2}{1.5cm}{\textbf{\romannumber{1}} {\tiny ( 
                \categoryonea\textsuperscript{A}, 
                \categoryoneb\textsuperscript{B}
        )}} \\ 
    \cline{4-4}
       &  &  & {\centering $z-fixed$ {\tiny(\numberconeonetwo)}} & \\ 
    \cline{4-5}
       &  &  & {\centering bad fit {\tiny(\numberconeonethree)}} & {\textbf{\romannumber{2}} {\tiny ( \categorytwo)}} \\ 
    \cline{3-5}
       &  & {\centering $\le3$ {\tiny(\numberconetwo)}} &   &  
        \textbf{\romannumber{3}} {\tiny ( 
                \categorythreea\textsuperscript{A}, 
                \categorythreeb\textsuperscript{B}
        )} \\  
    \cline{2-5}
       &  \multirow{2}{1.5cm}{\centering $2R_e > \SI{1.5}{\arcsecond}$ {\tiny(\numberctwo)}} & 
       $\ge 4$  {\tiny(\numberctwoone)}
       &  & 
        \textbf{\romannumber{4}} {\tiny (\categoryclear)}
        \\ 
    \cline{3-5}
       &  & $\le 3$  {\tiny(\numberctwotwo)} &  & \textbf{\romannumber{5}} {\tiny (\categorydiscard)} \\ 
    \bottomrule
\end{tabular} \\
\footnotesize{
\begin{flushleft}
A: \oii\ detected $\ge 2 \sigma$ level; \\ B: \oii\ not detected i.e. $< 2 \sigma$ level\\

 Detection Set: \romannumber{1}A, \romannumber{1}B, and \romannumber{3}A; Non-Detection Set: \romannumber{2}; `CLEAR' Subset: \romannumber{4}
\end{flushleft}}
\end{table}

\begin{figure}
    \centering
    \includegraphics[width=0.47\textwidth]{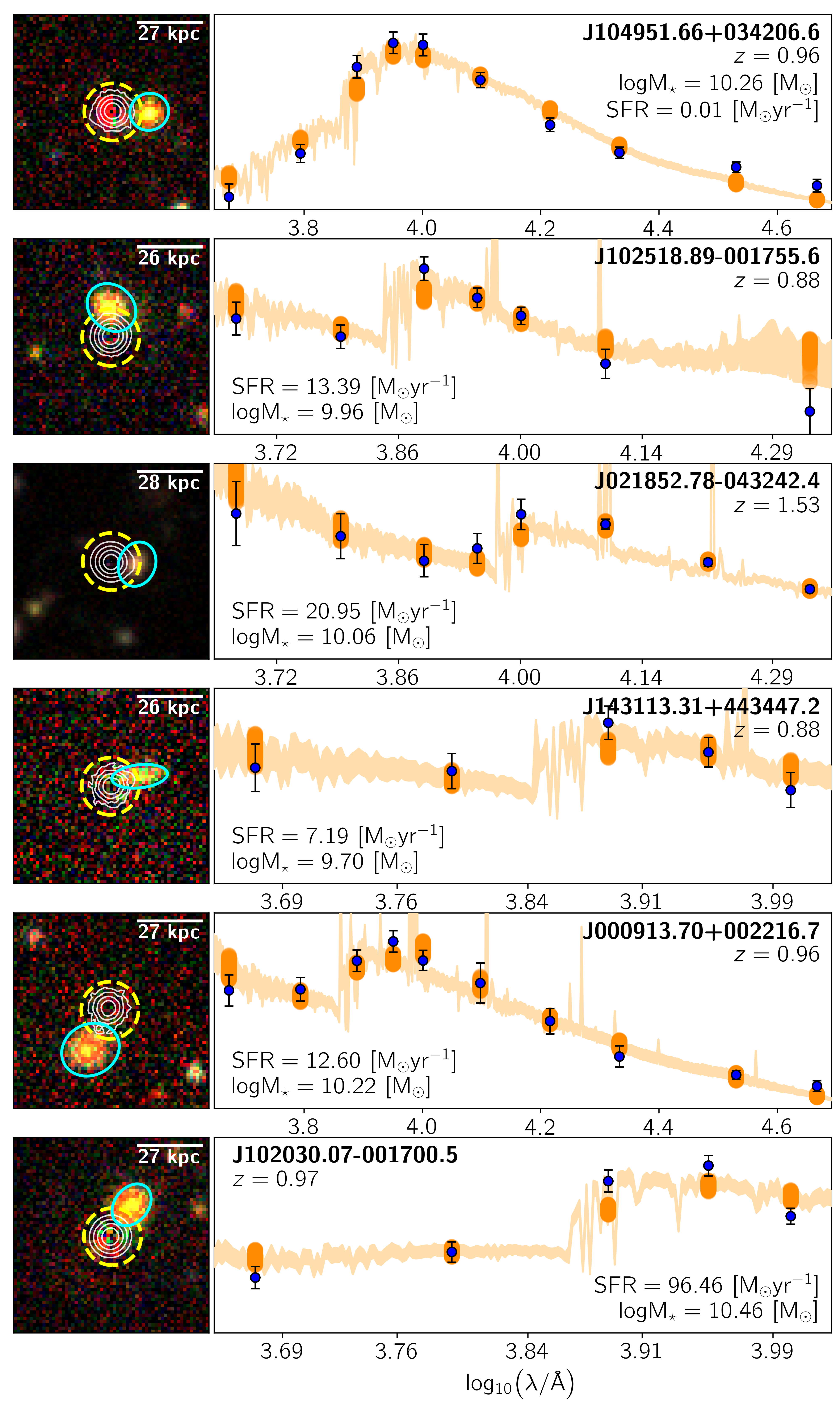}
    \caption{The postage stamp HSC color composite images, centered on the quasar, are depicted with gray contours. The SDSS fiber, with a radius of \SI{1.5}{\arcsecond}, is indicated by a dashed yellow circle, while the cyan-colored aperture highlights the USMgII host galaxy. The second column exhibits the corresponding multi-band best-fit SED model at the absorber redshift.}
    \label{fig:sedcollage}
\end{figure}

\subsection{Multi-band SED modelling}
\label{lab:sed}

In this section, we employ a multi-band spectral energy distribution (SED) model to investigate the absorber-galaxy connection based on photometric redshifts and to constrain the key physical parameters of USMgII galaxies. For a wide redshift range, $\allabszmin\ \le z \le \allabszmax$, of our USMgII galaxies, the filter set must be chosen to cover various key spectral features in SED modelling. For instance, Balmer/4000\AA\ break sensitive to the age and metallicity of stellar populations has shifted between  $r$ and $J$ band.  In addition, the UV excess particularly at wavelengths shorter than 2500\AA\ which is indicative of ongoing star formation activity is covered between $g$ and $i$-band. 

Besides the HSC optical passbands,  we also search for near-infrared, $J, H, K_s$, passband images from the VISTA Surveys\footnote{\href{http://vsa.roe.ac.uk/}{http://vsa.roe.ac.uk/}}. Among all the $\usmgiisightlinesclosebydetection$\ sightlines, we found $\closebydetectionvideosightlines$, $ \closebydetectionvikingsightlines$, and $\closebydetectionvhssightlines$ quasars covered in the VIDEO (at 5$\sigma$ depth of $J=24.5,~H=24.0,~K_s=23.5$), VIKING (at 5$\sigma$ depth of $J=22.1,~H=21.5,~K_s=21.2$) and VHS (at 5$\sigma$ depth of $J=21.2,~H=20.6,~K_s=20.0$) survey, respectively. For the near-IR flux measurement, we employed forced photometry using the Source Extractor based on the aperture from HSC images. We have corrected the HSC apertures to match the seeing of about \SI{0.8}{\arcsecond} offered by VISTA.  In the case of non-detection, we have estimated flux upper limits at 3$\sigma$ level. Furthermore, we also used the unWISE, $W1$, and  $W2$, fluxes for the sources along \closebydetectiondecamwisesightlines\ sightlines from Dark Energy Camera Legacy Survey \citep{Dey2019AJ....157..168D}. To obtain more secure results, we demand that the target should be detected (at $3\sigma$ level) in any of the four filters mentioned above, resulted in $\closebydetectionthreesigmaminfourfilter$ sources. \par

The multi-band photometric fluxes are modeled using BAGPIPES\footnote{\href{https://bagpipes.readthedocs.io/en/latest/}{https://bagpipes.readthedocs.io/en/latest/}} \citep{Carnall2018} utilizing the MultiNest sampling algorithm \citep{Feroz2009}. For this, we use a simple model considering a delayed star formation history with a wide parameter space for age between 50 Myr to 13.5 Gyr, mass formed ( $ 6 \le \rm log(M_*/M_{\odot}) \le 13$), and metallicity ($\rm 0.005 < [Z/H] < 5$). We assume the dust extinction law of \citet{Calzetti1994ApJ...429..582C} with total extinction $0 < A_v < 4$. \par

Firstly, we model the galaxies by varying the redshift over a range of $0.4 \le z \le 2.0$. Interestingly, despite the limited band measurement and large associated uncertainties, about \detectionphotozmatched\ out of \closebydetectionthreesigmaminfourfilter\ sources are found to match the absorber redshift within a typical photometric error of $\Delta z/(1 + z_{spec} ) \le 0.15$ \citep{Ilbert2006A&A...457..841I} (see, Fig~\ref{fig:sedcollage}). Recall that, apart from the direct detection of \oii\ nebular emission at $\ge 2\sigma$ level, the strong  \oii\ emission in the stacked spectrum signifies that most of our detections are legitimate USMgII host galaxies. Therefore, we further model the galaxies by fixing the redshift within $z_{abs} \pm 1000 \rm km\ s^{-1}$. We obtained a good fit for additional  \detectionzfixgoodfitadditional\ out of \closebydetectionthreesigmaminfourfilter\ absorbers, resulting in a total of \categoryonedetection\ absorber host (see column 4 of Table~\ref{tab:usmgiisummary}). In the case of multiple absorbers along a quasar sightline, if the photometric redshift matches with any other absorber, we drop the galaxy from our detection set (see, Table~\ref{tab:usmgiisummary}, Category \romannumber{2}). Using \categoryoneb, out of the above \categoryonedetection, absorbers with \oii $< 2 \sigma$ level and best photo-z estimates, we further generate the spectral stack to detect the \oii\ nebular emission. A color composite image for this subset is shown in rows 4 to 7 of Fig.~\ref{fig:collage}. We detect a \oii\ emission line at  $\categoryoneboiisnr\ \sigma$ level (see, Fig~\ref{fig:collage}, panel b). Finally, we consider all the \detectionoiicovered\ systems, with and without \oii\ detection, and detect a clear \oiiab\ doublet profile in the composite spectrum at the $\categoryoneaboiisnr\ \sigma$ level (see, Fig~\ref{fig:collage}, panel c). The higher detection fraction, along with the observed \oii\ emission, reaffirm that these are genuine USMgII galaxies.

{\it In summary, utilizing the  \oii\ nebular emission detected on top of individual quasar spectra we detect \detectionoiidetected\ USMgII host galaxies. Additionally, based on \oii\ emission in the stacked spectra, in conjunction with the photometric redshift estimates, we detect \detectionoiinondetectednotcovered\  potential absorber galaxies, resulting in a total detection of \detection\ USMgII galaxies along  \totalnumberofusmgiisightlines\ sightlines in deep HSC images. Among them,  \categorythreea\ absorbers are selected solely based on \oii\ nebular emission line due to limited HSC passbands. Hereinafter, the subset of \categoryonedetection\ galaxies with best fit SED model parameters is employed to study the galaxy properties in detail. The catalog is available online as supplementary material.}

\subsection{Galaxy geometry}
\label{sec:galgeo}
To infer the distribution of the metal contents and the gas dynamics in the circumgalactic medium of USMgII hosts, we first measure the azimuthal angle ($\phi$) of the quasar sightline with respect to the galaxy's major axis. For this, we model the galaxy with a s\'ersic profile and estimate the galaxy ellipticity ($e=1-\frac{b}{a}$), and position angle on deep HSC images \citep{Lundgren2021ApJ...913...50L}. Since $\phi$ measurement requires a robust estimate of galaxy position angle, we select the galaxies subset with higher ellipticity, $e > 0.2$, and consistent measurement of position angle in all the available HSC filters. Finally, visual scrutiny is carried out to remove the outliers, including any possible merger or cases with galaxy on top of quasars, and to generate a subsample for examining the dependence of absorber property on azimuthal angle. We have found \detectiongoodpa\ systems with a robust position angle measurement and high ellipticity $e > 0.2$.

\begin{figure}
    \centering
    \includegraphics[width=0.5\textwidth]{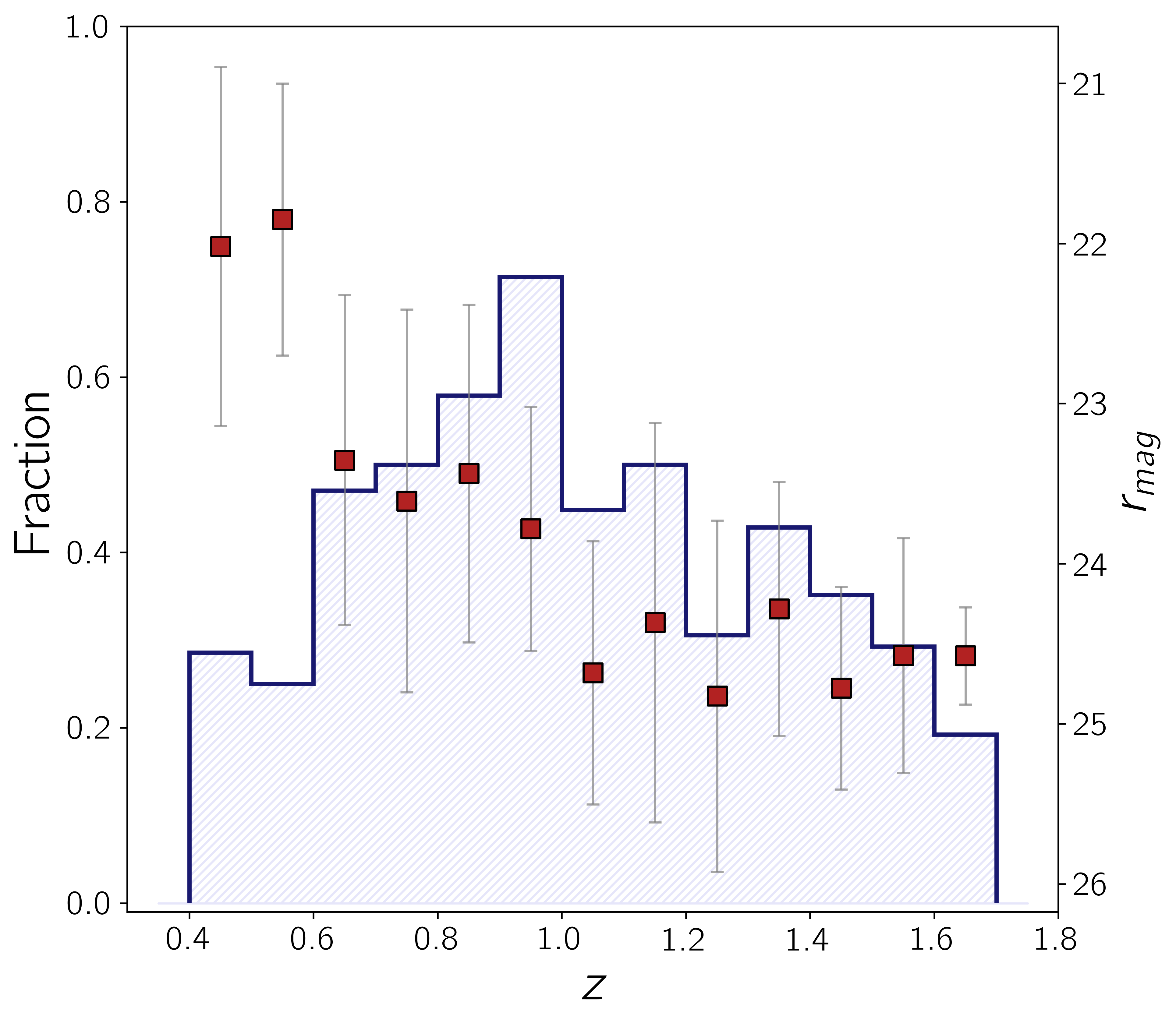}
    \caption{Distribution of the detection rate of USMgII absorber galaxies with redshift. The square shows the average $r-$band magnitude of galaxies per redshift bin, along with the 16th and 84th percentiles.}
    \label{fig:detectionrate}
\end{figure}

\begin{figure}
    \centering
  \includegraphics[width=0.49\textwidth]{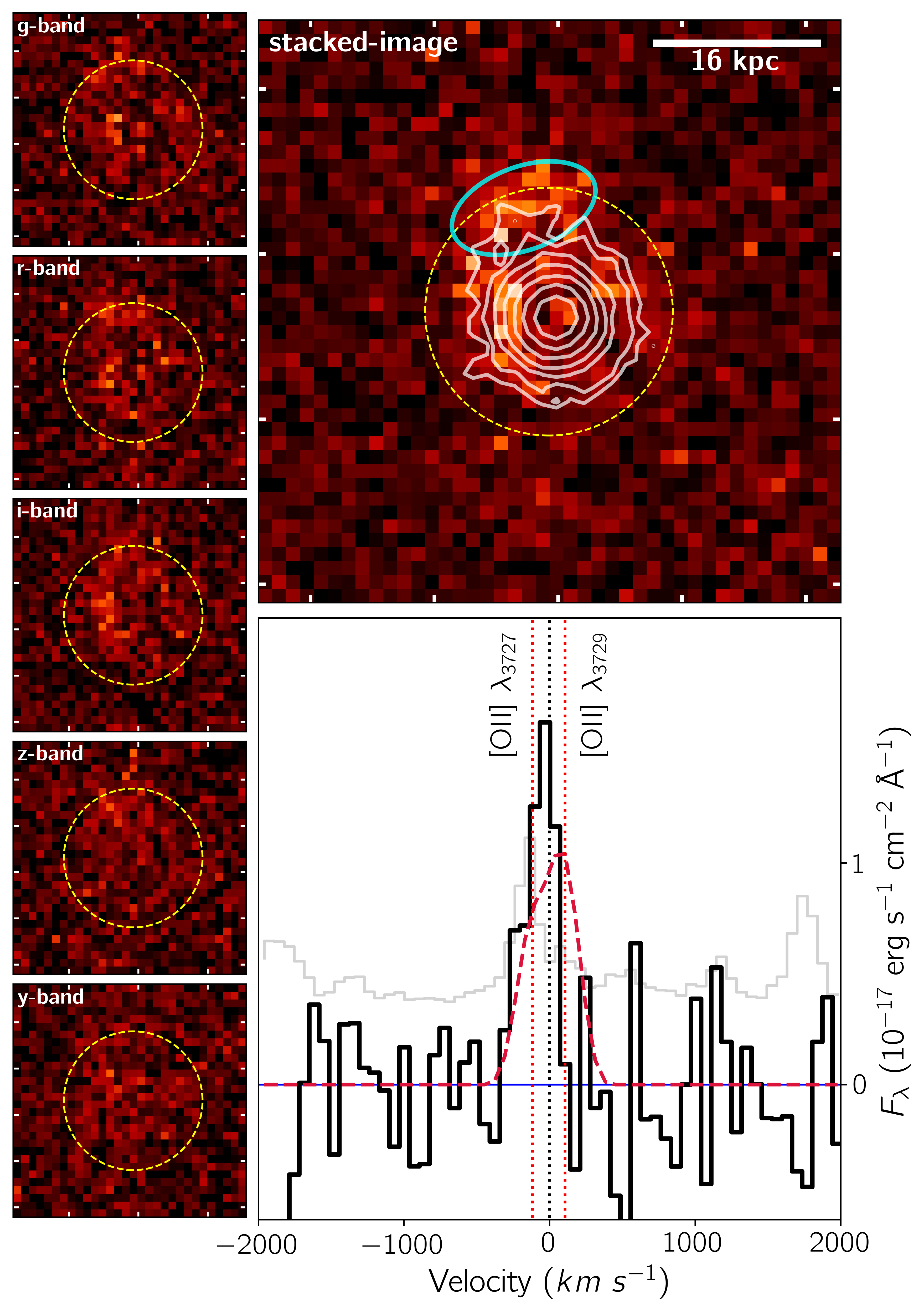}
    \caption{ A potential galaxy candidate from the faint end population at $z_{abs} = 1.018$, along  J234451.70+000603.2 quasar sightline at  $z_{qso} = 2.4$. {\it Left panel:} The postage stamp HSC multi-band images after quasar removal. {\it Top right panel:} HSC five band median coadded image showing a potential host galaxy detected at $2.7 \sigma$ level, marked as cyan aperture. The SDSS fiber, with a radius of \SI{1.5}{\arcsecond}, is indicated by a dashed circle whereas the quasar is shown as contours. {\it Bottom right panel:} An emission signature at the expected  \oiiab\ wavelength, detected at $\sim\darkexampleoiisigma\ \sigma$ level.}
    \label{fig:darkcollage}
\end{figure}

\section{Results}
\subsection{Detection rate of USMgII absorber galaxies} 
\label{lab:detectionfrac}
In the early efforts to map the host galaxies and environments of USMgII absorbers, \citet{Nestor2007ApJ...658..185N} utilized the deep optical images that revealed the existence of at least one bright galaxy with luminosity $\ge 0.3 L^{\star}$ within an impact parameter of 40 kpc at redshifts of $0.42 < z < 0.84$. Recent efforts using advanced integral field spectroscopy techniques have revealed that multiple galaxies may be associated with the strong \mgii absorber. 
In MusE GAs FLOw and Wind (MEGAFLOW) survey,  \citet{Schroetter2019MNRAS.490.4368S} studied the circumgalactic medium around z $\sim$1 star-forming galaxies and detected one or more galaxies for 59/79 (75\%) systems  with stellar masses ($M_{\star}$) from $10^9$ to $10^{11}\ M_{\sun}$.  \citet{Dutta2020MNRAS.499.5022D} have searched the absorber host galaxies with SFR and stellar mass down to 0.1 $\rm M_{\odot} yr^{-1}$ and $10^7 M_{\star}$, respectively, around quasar out to $\approx 200$ kpc. They observed a high detection rate of $78\%$, with 67\% \mgii absorbers associated with the non-isolated environments. \citet{Lundgren2021ApJ...913...50L} reported an exceptionally high detection rate of $\sim 89\%$ for strong \mgii systems in the deep, high-resolution 3D HST survey over an impact parameter of 200 kpc and SFR limit of $\rm > 1.3 M_{\odot} yr^{-1}$ at z = 1. Although the handful of efforts mentioned above provide the 3D view with high spatial resolution and sensitivity, have been limited and time-expensive, which largely missed the absorber counterparts at low-impact parameters. \citet{Guha2022MNRAS.513.3836G, Guha2024MNRAS.527.5075G} have targeted the bright galaxies with $m_r < 23.6$  for long-slit spectroscopy observations. They detect massive galaxies ($M_{\star} > 10^{10.2} M_{\odot}$, survey completeness of $0.3 L^{\star}$) with about 29\% of the systems likely associated with isolated galaxies while $\sim 21\%$ are in groups. However, in 50\% cases, no potential host is identified based on either spectroscopic or photometric redshift within 50 kpc.

Within our large sample of USMgII absorbers, probing stellar mass down to $10^{\minstellarmassdetected} M_{\odot}$ derived from BAGPIPES, we detect a galaxy up to 
\ipendkpc kpc (\SI{\alldetectionseparcsecmax}{\arcsecond}) for \detectionpercent\% of the cases, where the $2R_e$ falls within the fiber radius of \SI{1.5}{\arcsecond}. We obtain a detection rate of \detectionpercentminfourfilter\% by stipulating an availability of at least 4 HSC passbands to detect the galaxy. The detection rate as a function of redshift is shown in Figure~\ref{fig:detectionrate}, which remains broadly consistent at 30-40\%. For these USMgII systems the direct \oii\ detection rate, at $2 \sigma$ level, in the SDSS fiber spectrum is \oiidetectiontwosigmapercent\% which raises to $\sim \oiidetectiononepfivesigmapercent\%$ if we consider the detection threshold of $\sim1.5\sigma$ level. This further aligns with the expectations based on the observed strong \oii\ nebular emission in the stacked spectra \citep{Noterdaeme2010MNRAS.403..906N, Menard2011MNRAS.417..801M,Joshi2018MNRAS.476..210J} as well as observed anti-correlation between \ew\ vs impact parameter \citep{Chen2010ApJ...714.1521C, Nielsen2013ApJ...776..115N}.

Furthermore, we note that for \nondetectionpercent\% of absorbers, no counterpart is detected in individual filters at $3 \sigma$ level or even in the stacked images. Whereas for \hintproximdetectionpercent\% of cases, a nearby faint source is seen in the stacked images.  It suggests that a handful of USMgII systems are either hosted by dusty star-forming galaxies, post-starburst galaxies, or very faint/dwarf galaxies. It is pertinent to recall that we detected \oii\ emission line signature along the clear sightlines, indicating the possible presence of faint galaxies. Therefore, such systems are also potential candidates for the dark galaxies \citep{Matsuoka2012AJ....144..159M, Straka2013MNRAS.436.3200S}, and further explored in Das et al., in preparation. One such example for the quasar sightline J234451.70$+$000603.2 is shown in Figure~\ref{fig:darkcollage} where \oii\ line signature is detected at the $\darkexampleoiisigma\ \sigma$ level whereas a potential host is only discernible in the stacked image. Further, spectroscopic and deep imaging follow-up of these cases will help to confirm and study the origin of USMgII absorbers in faint galaxy population.

\begin{figure}
    \centering
    \includegraphics[width=0.49\textwidth]{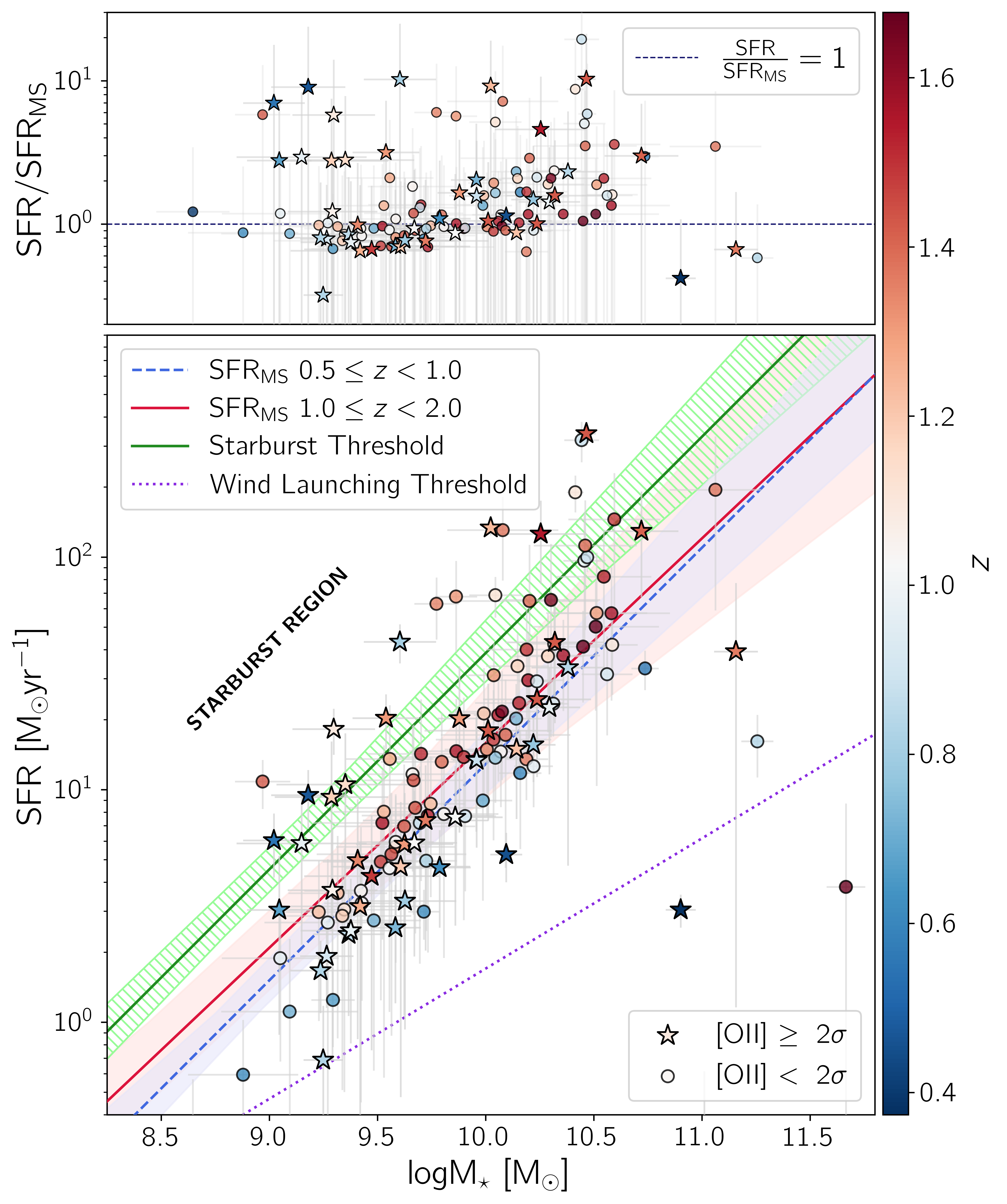}
    \caption{Main sequence, SFR versus stellar mass, USMgII absorber host galaxies. The {\it dashed-blue} and {\it solid-red} lines depict the best linear fit for main sequence galaxies at $ 0.5 \le z < 1$  and $1 \le z < 2$, respectively while the {\it solid-green} line represents the sequence for starburst galaxies from \citet{Bisigello2018A&A...609A..82B}. The symbols are color-coded with the redshift. The {\it shaded} and {\it hatched} regions indicate the $1\sigma$ confidence interval. Additionally, the {\it dotted-line} represents the SFR threshold for wind launching from \citet{Murray2011ApJ...735...66M}. {\it Top panel:} SFR of USMgII galaxies normalized by the main sequence star-forming galaxy at respective redshift.}
    \label{fig:SFR}
\end{figure}

\subsection{SFR in USMgII host galaxies}
\label{lab:SFR}
Evidenced by the SFR and \ew\ correlation, the USMgII galaxies are often found to be likely starburst in nature. \citet{Bouche2007ApJ...669L...5B} have observed 21 very-strong MgII absorbers with \ew $> 2$\AA\ at $z \sim 1$, using VLT/SINFONI near-infrared IFU, and detected the majority of absorber host galaxies with  $\rho$ ranging from 0.2 to 4 arcsec (i.e., $\rho\ \le $ 25~kpc) with SFR (H$\alpha$) between 1-20  $\rm M_{\odot}\ yr^{-1}$. Taking advantage of the large 3D view offered by MUSE/VLT, the MEGAFLOW survey has studied 9 systems with strong  \ew $> 2$\AA\ over $ 0.8 < z < 1.6$. They found all, but one, absorber galaxy having star formation rates greater than $2.5 \rm M_{\odot}\ yr^{-1}$, with the average impact parameter of 22 kpc. In long-slit observations of the galaxies near the quasar, \citet{Guha2022MNRAS.513.3836G, Guha2024MNRAS.527.5075G} have targeted 63 USMgII absorbers and found a high detection rate of $\sim 50$\% at average $\rho \sim$ 19kpc. The stellar mass for their USMgII galaxies ranges between  $\rm 10.2 \le log M_{\star} [M_{\odot} ] \le 12.01$ with an average $\rm log M_{\star} = 10.73 M_{\odot}$. These galaxies are found to have an average SFR of $\sim 21.16\ \rm M_{\odot}\ yr^{-1}$.

Note that, in the SDSS and BOSS finite fiber radius of \SI{1.5}{\arcsecond}, and \SI{1}{\arcsecond}, respectively, only a part of the galaxy comes into the fiber, as a result, our measured SFR inferred from \oii\ nebular emission is a lower limit. Among \closebydetectionoiidetected\ USMgII galaxies with direct \oii\ detection, for 15 galaxies the SFR based on SED is consistent with \oii\ based SFR. Hence, here we relied on SED model-based SFR measurements. All the USMgII galaxies having  SFR ranging between \minsfrdetected\ to $\maxsfrdetected\ \rm M_{\odot}\ yr^{-1}$ with an average SFR in USMgII galaxies is found to be \meansfrdetected $\rm M_{\odot}\ yr^{-1}$. \par

In general, the metal absorber properties, such as metallicity, the strength of \ew\, are interlinked with the galaxy properties, such as $M_{\star}$, and SFR \citep{Lan2014ApJ...795...31L, Rubin2018ApJ...853...95R, Dutta2020MNRAS.499.5022D}. In addition, a color dependence is also observed, showing that the star-forming galaxies host stronger absorbers than the passive ones (\citealt{Lan2018ApJ...866...36L, Bordoloi2011ApJ...743...10B}, but see also, \citealt{Chen2010ApJ...714.1521C}). To examine the dependence of USMgII metal absorption on the host galaxy properties, we have calculated Kendall's $\tau$-coefficient of \ew\ with $M_{\star}$, and found a weak positive correlation with $\tau_k = \tauewvsmstarstat$ and null probability of $p_{k} = \tauewvsmstarpvalue$. The \ew\ is also correlated with SFR  ($\tau_k = \tauewvssfrstat$  and $p_{k} = \tauewvssfrpvalue$), albeit weaker than the correlation found for \ew\ and $\rm M_{\star}$. As the metal enrichment of the CGM is the result of supernova, the ratio of $\alpha$-elements to Fe should reflect the supernova rates, and the efficiency of outflows in enriching the CGM. Therefore, we study the link of $M_{\star}$ and SFR with the equivalent width of \feiia (\ewfeii), tracing $10^4K$ gas, and the equivalent width ratio between \mgiia\ and \feiia ($R =$ \ew/\ewfeii). Both the \ewfeii\ ($\tau_k = \tauewfeiivssfrstat$, $p_{k} =\tauewfeiivssfrpvalue$) and $R$ parameter ($\tau_k =\tauewfeiimgiivssfrstat$, $p_{k} = \tauewfeiimgiivssfrpvalue$) mildly correlated with SFR, whereas no such dependence is seen with stellar mass. This likely highlights the role of supernova-driven galactic outflows in the metal enrichment of CGM. Interestingly, the USMgII absorbers along the `clear' sightlines show a relatively lower median $R = \averagefeiibymgiinondetection \pm \stdfeiibymgiinondetection$ compared to the `detection' set having $R = \averagefeiibymgiidetection \pm \stdfeiibymgiidetection$.  A two-sided Kolmogorov–Smirnov test finds a significant difference based on the $R$ parameter between the `clear' and `detection' subsets with a null probability of being drawn from the same parent distribution to be $P_{KS} = \kstestfeiibymgiipvalue$. As both  \mgii and \feii\ absorption are likely saturated at SDSS resolution, the difference in $R$ may likely related to the different kinematic spread of the gas which is essentially driven by the SFRs. This also supports the previous findings where \mgii systems with strong \ewfeii\ are found to have strong  \oii\ luminosity. Such a high $R$ is likely to arise in high metallicity sub-DLAs or high N(H{\sc  I}) DLAs \citep{Srianand1996ApJ...462..643S} and trace the \mgii absorber hosts with higher star formation \citep{Joshi2017MNRAS.471.1910J,Joshi2018MNRAS.476..210J}. 

In Figure~\ref{fig:SFR}, we compare the SFR of USMgII galaxies versus stellar mass along with the best fit relation for main sequence star-forming galaxies at $0.5 \le z \le 1$ and $1 \le z \le 2$, from  \citet{Bisigello2018A&A...609A..82B}. The points are color-coded with the absorber redshift.  Our USMgII absorbers span over a stellar mass of $\minstellarmassdetected\ \le \rm log~M_{\star} [M_{\odot}] \le  \maxstellarmassdetected\  $, with an average $\rm log M_{\star}$ of \averagestellarmass\ $\rm M_{\odot}$. A noticeable rise in SFR with stellar mass is evident in Figure~\ref{fig:SFR}. The subsets below and above the average $\rm log~M_{\star}$ have median SFR of \mediansfrbelowaveragestellarmass\ ($\pm \stdsfrbelowaveragestellarmassdex$ dex) and \mediansfraboveaveragestellarmass\ ($\pm\stdsfraboveaveragestellarmassdex$ dex) $\rm M_{\odot} yr^{-1}$, respectively.  In the top panel of Figure~\ref{fig:SFR}, we compare the SFR, normalized by the main sequence star-forming galaxies of similar mass and redshift, obtained from the best-fit relation by \citet{Popesso2023MNRAS.519.1526P}.  It is worth noting that star-forming galaxies often face significant dust obscuration, resulting in an underpredicted SFR by a factor of 2-3 in SED modeling routines  \citep{Wuyts2009ApJ...696..348W}. It is clear from Figure~\ref{fig:SFR} that USMgII absorber host galaxies are broadly consistent with the main sequence SF galaxies. Interestingly, around \starburstpercent\% of galaxies exhibit a starburst nature, characterized by star formation rates exceeding three times that of main-sequence galaxies  \citep{Elbaz2018A&A...616A.110E}. Furthermore,  the USMgII galaxies meet the threshold star formation rate for launching strong outflows \citep{Murray2011ApJ...735...66M}, shown as {\it dotted} lines in Figure~\ref{fig:SFR}.

\begin{figure}
    \centering
    \includegraphics[width=0.49\textwidth]{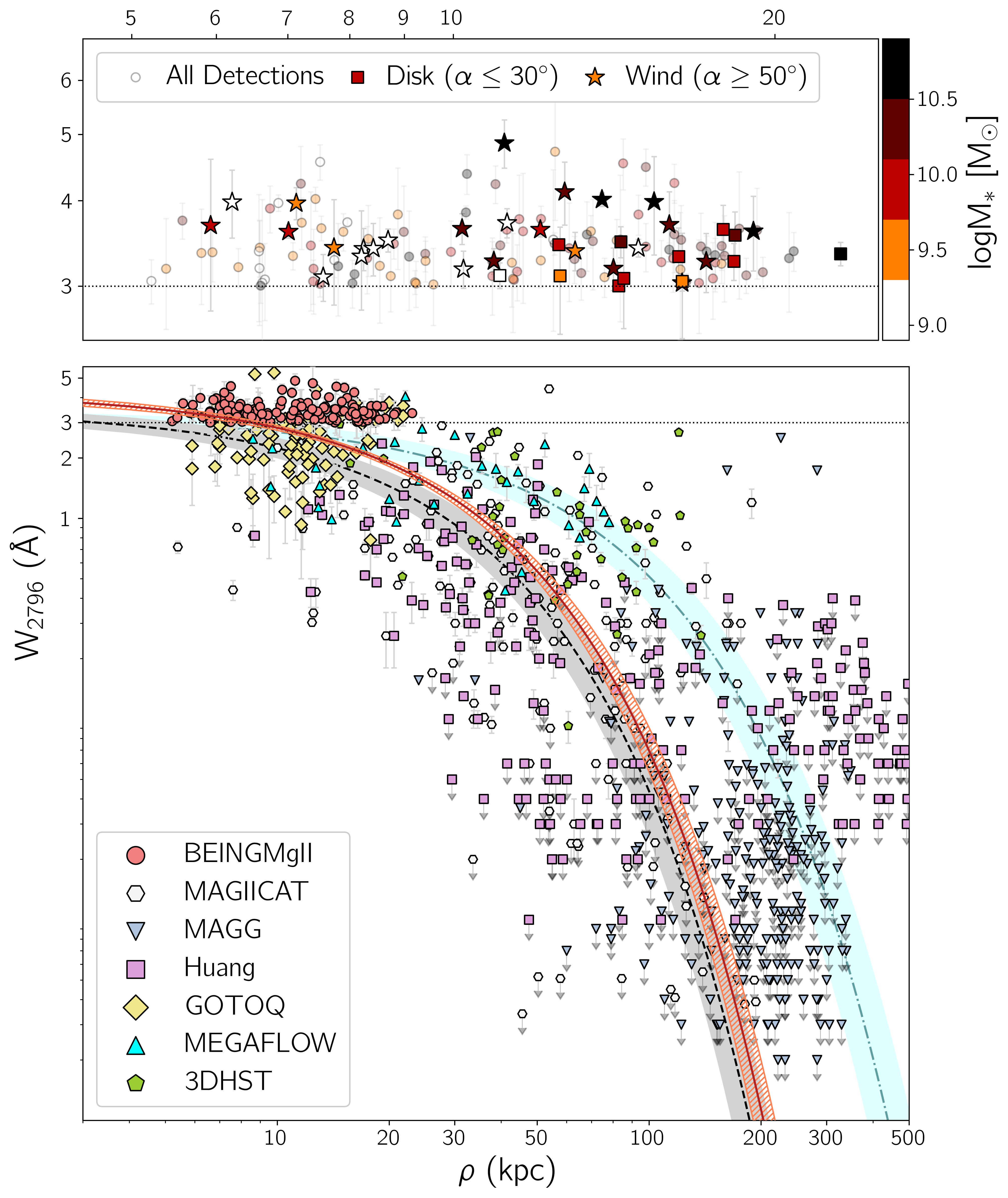}
    \caption{{\it Bottom panel:} Equivalent width (\ew) versus impact parameter ($\rho$) relation for MgII galaxies. The best-fit anti-correlation between \ew\ and $\rho$ is shown as a solid line along with the 16th and 84th percentiles indicated as a hatched region. The best-fitting relationships from the literature
    for $z \sim 1$ \mgii systems from \citet{Lundgren2021ApJ...913...50L} ({\it dot-dashed} line) and a global fit for $0 < z < 1.5$ from \citet{Guha2024MNRAS.527.5075G} ({\it dashed} line) are also plotted for reference. {\it Top panel:} Similar to the above for only USMgII galaxies ({\it circle}), color coded with the $M_{\star}$. The  quasar-galaxy pairs classified as `wind' with $\alpha \ge 50^{\circ}$ and `disk' with $\alpha \le 30^{\circ}$, are represented as {\it star} and {\it square} symbols, respectively. The wind subset preferentially shows higher \ew\ and large scatter relative to the disk subset.
    }
    \label{fig:EWD}
\end{figure}

\begin{figure*}
    \centering
    \includegraphics[width=0.45\textwidth]{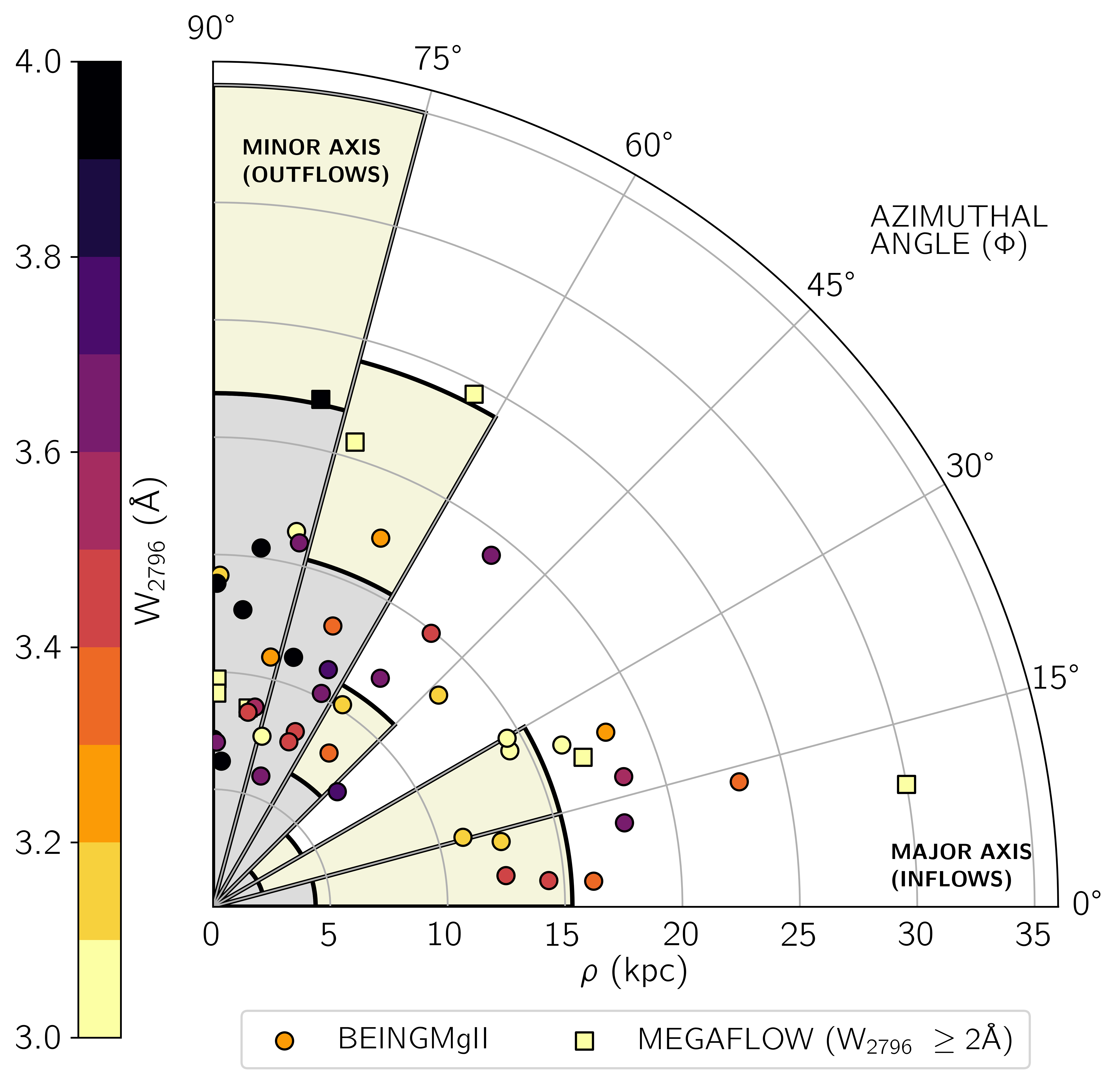}
    \includegraphics[width=0.45\textwidth]{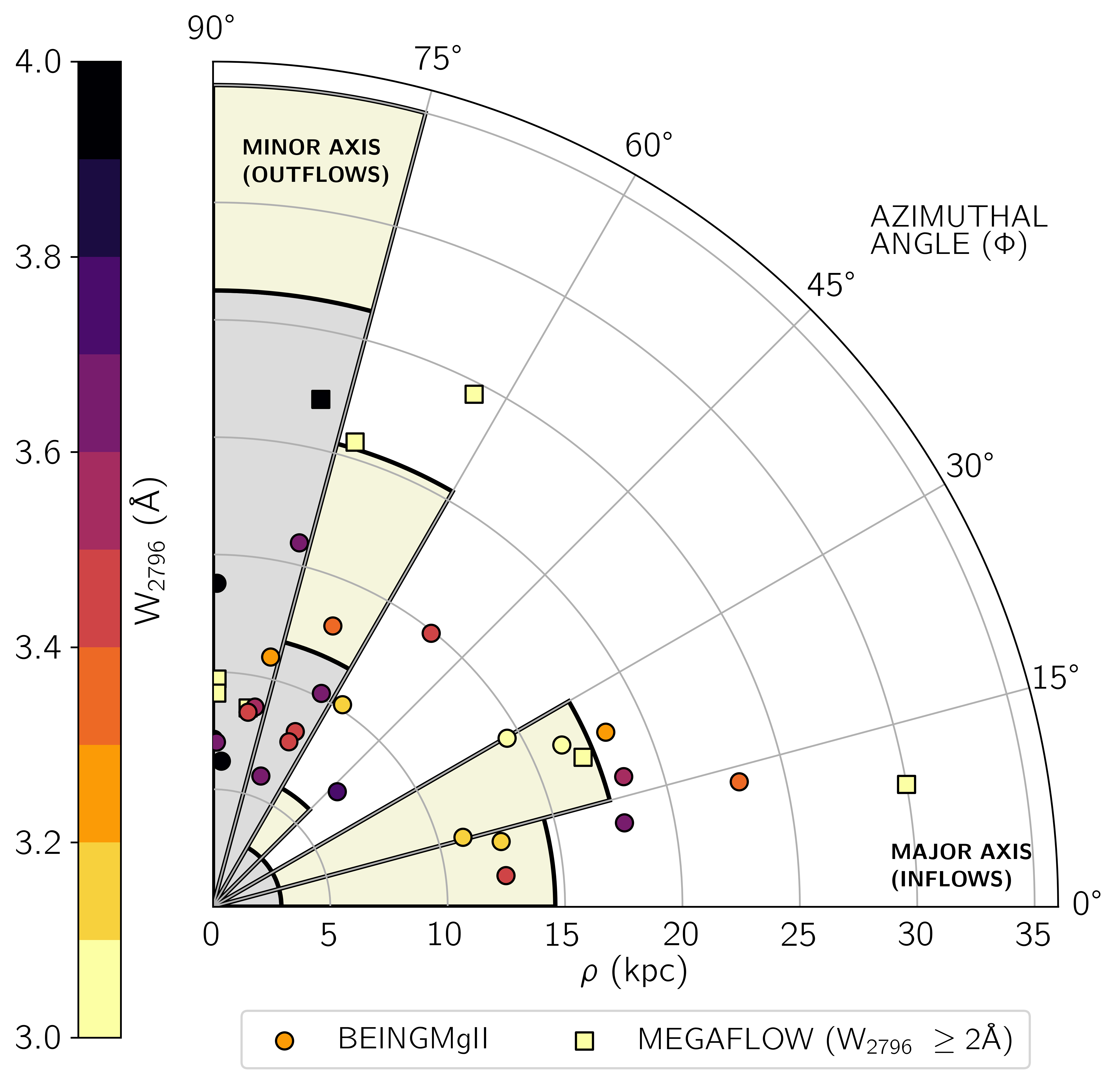}
    \caption{{\it Left Panel:} Distribution of \mgii absorbers as a function of azimuthal angle for the USMgII host galaxy with ellipticity $e \ge 0.2$. The symbol size represents the \ew/\ewfeii\ with the color denoting the strength of equivalent width (\ew). {\it Right Panel:} same as left  for the subset with ellipticity of $e \ge 0.3$. The histogram shows the cumulative $\phi$ distribution of the USMgII absorber, below (grey region) and above (yellow region) the median impact parameter of $\ipmediankpc$ kpc, respectively.}
    \label{fig:polar}
\end{figure*}

\subsection{ Equivalent width versus impact parameter}
\label{lab:ew}
Numerous attempts to establish the connection between absorber and galaxies have revealed an anti-correlation between  \ew\ and the impact parameter ($\rho$). It is interpreted as a decrease in the fraction of low-ionization metal-enriched gas with increasing distance from the galaxy core \citep{Chen2010ApJ...714.1521C, Nielsen2013ApJ...776..115N}. Additionally, the significant dispersion observed in the \ew\ vs $\rho$ relation is influenced by factors such as galaxy stellar mass \citep{Chen2010ApJ...714.1521C, Dutta2020MNRAS.499.5022D}, host inclination angle \citep{Kacprzak2011MNRAS.416.3118K},  and the interplay of various physical mechanisms such as gas in-outflow \citep{Bouche2012MNRAS.426..801B}. Recently, using an absorption selected sample, \citet{Lundgren2021ApJ...913...50L} have shown a significant evolution of this anti-correlation across a range of redshifts from $z\sim0.4$ to $z\sim1.5$ \citep[but see also,][]{Dutta2020MNRAS.499.5022D}. Additionally, \citet{Lan2020ApJ...897...97L} reported a strong co-evolution of gas covering fraction around star-forming galaxies over the redshift range from $z\sim0.4$ to $z\sim1.3$, with a strong dependency on galaxy stellar mass. 

The \ew\ vs $\rho$ relation for all the \detection\ USMgII hosts detected in this work is presented in Figure \ref{fig:EWD}. The $\rho$ spans from $\sim \ipstartkpc$ kpc to $\sim \ipendkpc$ kpc, with a median $\rho$ of $\sim \ipmediankpc$ kpc. The above $\rho$ for USMgII hosts are much larger than the predicted, i.e., $\rho \lesssim 5$kpc by the \ew-$\rho$ anti-correlation at $z \sim 0.5$ \citep{Nielsen2013ApJ...776..115N} and at $z ~\sim 1.0$~\citep{Lundgren2021ApJ...913...50L}, respectively.  Recently, \citet{Guha2022MNRAS.513.3836G, Guha2024MNRAS.527.5075G} have reported a broad impact parameter ranging between $6.3$ kpc to $120$ kpc for 38 USMgII absorber  host galaxies between $z\sim0.4$ to $0.8$. However, it is important to note that instead of originating at such large impact parameters, the USMgII absorbers could be due to unseen galaxies (see Section~\ref{lab:detectionfrac}). 

We model the \ew\ versus impact parameter using a log-linear model as, log\ew\ (\AA) = $\alpha$ + $\beta \times \rho$ (kpc), with a likelihood function given in equation 7 of
\citet[][]{Chen2010ApJ...714.1521C}, see also  \citet{Dutta2020MNRAS.499.5022D}, and sample the posterior probability density function using PyMultiNest. For this, we have included the \ew\ and impact parameter measurements from literature, including the MAGIICAT sample \citep{Nielsen2013ApJ...776..115N}, MEGAFLOW survey \citep{Schroetter2019MNRAS.490.4368S}, MAGG survey \citep{Dutta2020MNRAS.499.5022D},  \citet{Huang2021MNRAS.502.4743H}, 3D-HST \citep{Lundgren2021ApJ...913...50L}, and  \citet{Guha2023MNRAS.519.3319G} and obtained a best-fit parameter of $ \alpha = \ewdoverallfitbestfitalpha^{+ \ewdoverallfitbestfitalphaupper}_{- \ewdoverallfitbestfitalphalower}, \ \beta = \ewdoverallfitbestfitbeta^{+ \ewdoverallfitbestfitbetaupper}_{- \ewdoverallfitbestfitbetalower}$. In the case of multiple galaxies associated with an absorber, we considered the galaxy with the smallest impact parameter. Figure~\ref{fig:EWD} shows the best-fit log-linear model, along with 1$\sigma$ uncertainty in the shaded region. It is clear from the figure that the USMgII systems deviate from the best-fit relation at relatively larger impact parameters of $\gtrsim 15$ kpc  \citep[see also, ][]{Guha2024MNRAS.527.5075G}. \par

The large scatter observed in \ew\ as a function of $\rho$ in Figure \ref{fig:EWD} can be reconciled with the various routes contributing to their origin (see Section~\ref{sec:intro}). One possible way to address this is to look for the gas origin in galactic winds,  seen as expanding biconical flow perpendicular to the galaxy disc \citep{Nelson2019MNRAS.490.3234N}, and gas accretion traced as co-rotating gas along the major axis \citep{Zabl2021MNRAS.507.4294Z}.  In most cases among our USMgII hosts, only one galaxy is observed around the quasar, providing an opportunity to test whether the absorption originates from outflowing winds or gas inflow. For this, we divided the sample into two subsets, i.e., closer to the minor ($\phi >  50^\circ$) axis -``{\it Wind} subset", likely to trace the outflowing wind ({\it blue circle}) and major ($\phi < 30^\circ$) axis - ``{\it Disk} subset", likely to trace gas inflows ({\it red circle}) [see top panel of Figure~\ref{fig:EWD}]. The above two bins consist of a total of \totaldiskwindsource\ systems, with \totaldisksource\ along the major and \totalwindsource\ along the minor axis.  It is clear from Figure~\ref{fig:EWD} that the strongest USMgII systems in our sample preferentially belong to the {\it Wind}-subset. The average \ew\ of the USMgII absorbers in {\it Wind}-subset is \windsubsetmedianew $\pm$\windsubsetstdew\AA\, whereas the {\it Disk}-subset has a slightly lower average \ew\ of \disksubsetmedianew $\pm$\disksubsetstdew\AA. It is evident that, in comparison to {\it Disk}-subset the scatter in \ew\ is governed by the {\it Wind}-subset.

Moreover, a near-constant \ew\ of the wind subset up to large impact parameters is consistent with the high-resolution cosmological simulations where the galactic winds substantially enhance the amount of cold gas in the halo, as evidenced by high covering fractions of H{~\sc i} and \mgii out to the virial radius and beyond \citep[see,][]{Suresh2019MNRAS.483.4040S}. On the other hand, the {\it Disk}-subset shows a relatively smaller scatter, suggests that the interplay between various physical origins, e.g., extended gaseous disks and galactic winds, collectively contributes to the significant dispersion in  \ew-$\rho$ relationship \citep[see also,][]{Bouche2012MNRAS.426..801B}. 
Finally, we assess for any dependence of  \ew-$\rho$ relation with redshift. We find no significant evolution for USMgII systems over a broad redshift range ($\allabszmin\ \le z \le \allabszmax$). This is in agreement with \citet{Guha2024MNRAS.527.5075G} where no significant redshift evolution is seen for USMgII absorbers within an impact parameter of $\le 120$ kpc. \par

\subsection{Gas distribution: azimuthal angle dependence}
\label{lab:alpha}
Earlier attempts to map the distribution of CGM gas, involving observations of galaxy samples and stacking experiments, have shown a preference for additional absorption to occur along both the major and minor axes of galaxies up to distances of 50 kpc \citep{Bordoloi2011ApJ...743...10B, Kacprzak2012ApJ...760L...7K, Lan2018ApJ...866...36L, Schroetter2019MNRAS.490.4368S, Zabl2019MNRAS.485.1961Z}. Moreover, utilizing redshift z $\sim$1 galaxies from the 3D HST survey, \citet{Lundgren2021ApJ...913...50L}  identified that this preference towards the semi-minor axis persists up to around $\sim$ 80 kpc, signifying the extent of galactic outflows. A strong dependence of \mgii emission as a function of inclination of the central galaxy, where edge-on galaxies clearly show enhanced emission along
the minor axis, while face-on galaxies showing weaker and isotropic emission supports biconical outflowing geometry 
 perpendicular to the galactic disk \citep{Guo2023Natur.624...53G}. The tendency for strong absorbers to cluster near the minor axis is commonly associated with starburst-driven outflows, while the excess absorption near the galaxy's major axis is attributed to gas accretion. Such an-isotropic gas distribution, with higher average metallicity along the minor versus major axes of galaxies, is also seen in cosmological hydrodynamical simulations \citep{Peroux2020MNRAS.499.2462P}.

Utilizing our dataset of USMgII absorbers, which effectively maps the CGM at low-impact parameters where gas flows are more pronounced, we examine their azimuthal angle dependence. 
Here, we additionally incorporate \megaflowabsorbercount\ strong absorbers (\ew $\ge 2$\AA), including a USMgII system found in the {\it Wind}-subset,
from the MEGAFLOW survey \citep[][]{Zabl2019MNRAS.485.1961Z, Zabl2021MNRAS.507.4294Z}, characterized by low impact parameters and reliable galaxy morphology measurements. In Figure~\ref{fig:polar}, we show the azimuthal angle distribution for the ``Wind'' and ``Disk'' subset, comprising of a total of \totaldiskwindsource\ systems with \totaldisksource\ along the major and \totalwindsource\ along the minor axis (see Section~\ref{lab:ew}). It is clear from Figure~\ref{fig:polar} that the strongest USMgII systems in our sample are preferentially observed along the direction of galactic winds.  For instance, among  \detectiongoodpaewfourangstrom\ absorbers with \ew$>$4\AA\ none of the absorbers are observed along the major axis. The KS-test rejects the null hypothesis of the two distributions being drawn from the same parent population at $P_{KS} =$\diskwindewpvalue. At first glance in Figure~\ref{fig:polar}, there appears to be an evident preference for azimuthal angle. To quantify this further, we divide the sample into three equal azimuthal angle bins of 30$^{\circ}$ each.  The $\phi$ bins of $0^{\circ}-30^{\circ}$ and $60^{\circ}-90^{\circ}$, i.e. close to the direction of the major and minor axis, comprise \detectiongoodpamajorpercent\% (\detectiongoodpamajor) and \detectiongoodpaminorpercent\% (\detectiongoodpaminor) out of total \detectiongoodpa\ USMgII systems, respectively. While the lowest average number of absorbers, i.e., \detectiongoodpainbetweenmajorminorpercent\% (\detectiongoodpainbetweenmajorminor) systems are observed along the  $\phi \sim 30^{\circ}-60^{\circ}$, hints at a bimodality in the absorber distribution. 

Using the high-resolution TNG50 cosmological simulation, \citet{DeFelippis2021ApJ...923...56D} have compared the CGM of $z \sim 1$ star-forming galaxies with strong Mg II absorbers (\ew $> 0.5$\AA) from the MEGAFLOW survey.  Contrary to the result from observations, they showed that Mg II absorption along the minor axis is weaker than along the major axis, with no net Mg II outflows along the minor axis \citep[see also,][]{Ho2020ApJ...904...76H}. In addition, they show that the azimuthal dependence of absorbers is very sensitive to the inclination angle of the sight line.  Therefore, we test the azimuthal angle dependence with a stringent ellipticity of $e > 0.3$. A striking bimodal distribution of USMgII absorbers is evident, with the majority of absorbers seen within 30$^{\circ}$ of the major and minor axis, (see Figure~\ref{fig:polar}). This implies that the cool gas traced by ultra-strong \mgii absorbers likely emanates from outflows propelled along the minor axis and IGM gas accretion in conjunction with the fountain flows, ISM, and satellite galaxies leading to the wide velocity spread along the major axis \citep{Guo2023Natur.624...53G,Rubin2022ApJ...936..171R,Perrotta2024arXiv240910013P}.

We further investigated the $\phi$ dependence on galaxy properties governing the outflows, e.g. sSFR, and \feii/\mgii abundance ratio ($R$). The average sSFR along the wind and inflow direction is found to be $\rm log(sSFR/yr^{-1}) = \averagessfrdiskwind$ which indicates that USMgII systems are hosted in galaxies with high sSFR. The $R$ is also found to be similar, with an average $R =$ \averagefeiibymgiidiskwind, which indicates that at average impact parameters of $\ipmediankpc$ kpc traced in our sample, the gas along the wind and disk plane is highly metal enriched.

\begin{figure}
    \centering
    \includegraphics[width=0.45\textwidth]{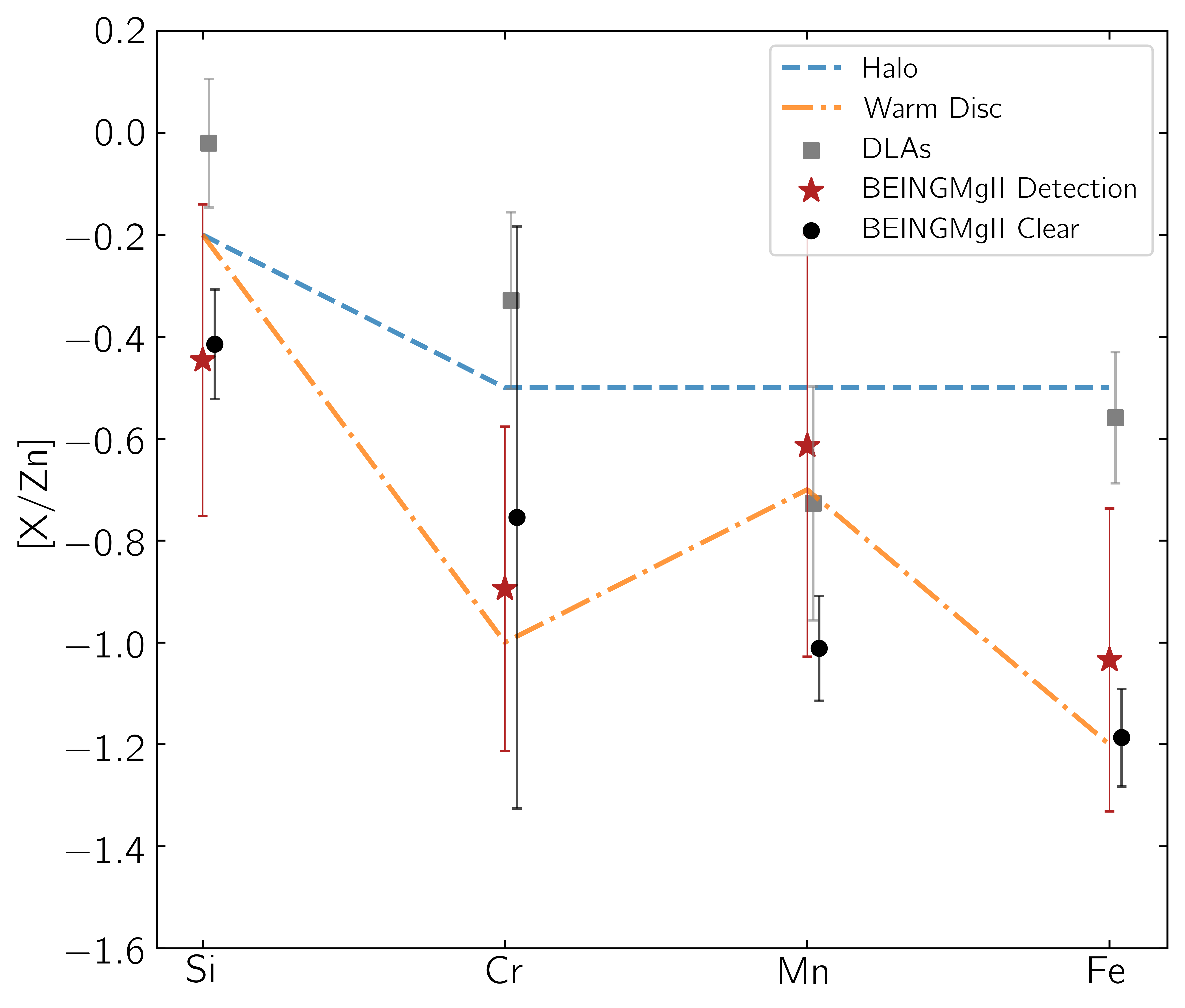}
    \caption{The relative abundances, [X/Zn], along detection versus clear sightlines, for several species compared with the depletion patterns of the
Milky Way halo (dashed line) and warm Disc (dashed-dot line) gas from \citet{Welty1999ApJ...512..636W}. The zero point of the ordinate corresponds to the solar metallicity. } 
    \label{fig:ABUNDANCE}
\end{figure}

\subsection{USMgII gas properties along detection versus 'clear' sightlines}
Strong \mgii systems generally trace a large amount of neutral gas.  In a low-redshift ($z < 1.65$)  study of \mgii absorbers with \ew $> 0.3$\AA, \citet{Rao2006ApJ...636..610R} found that all the DLA  systems (log N(H{\sc i}) $\rm [cm^{-2}] > 20.3$) have \ew $> 0.6$\AA. This suggests that most of our USMgII systems may likely be DLAs. To explore this further, we compare the metal content in USMgII absorbing gas along the detection set tracing the galaxy halo and the clear sightlines, with no galaxy detected within 20 kpc. To derive the typical metal content, we produced a median stacked spectrum by using the spectral region redward of Ly$\alpha$ forest, for both the detection and clear subset, consisting of \detection\ and \usmgiiclearsightlinesabsorbers\ absorbers, respectively. We detect high-ionization lines (Si{~\sc iv}, C{~\sc iv}), low-ionization lines (Fe~{\sc ii}, Si~{\sc ii}, Zn~{\sc ii}, Cr~{\sc ii}, Mg~{\sc i}), and weak transition lines (Fe~{\sc ii} $\lambda\lambda$2249,2260, Ni~{\sc ii}). We model the metal lines as a single Gaussian profile to measure the rest-frame equivalent widths. The rest equivalent widths of Zn~{\sc II} $\lambda\lambda$2026,2062\AA, is estimated by removing the blending with Mg~{\sc i} $\lambda2026$ and Cr~{\sc ii} $\lambda$2062  as described by \citet{York2006MNRAS.367..945Y}.
It is worth noting that despite the small apparent optical depth, most of the lines are saturated due to the low spectral resolution of SDSS ($R \sim 2000$). We estimate the column density for various species using the curve of growth of different absorption lines that originate with a broad range of oscillator strengths \citep{Jenkins1986ApJ...304..739J,Noterdaeme2014A&A...566A..24N,Joshi2017MNRAS.465..701J}.

In Figure~\ref{fig:ABUNDANCE}, we show the relative abundance ratio between refractory element (Si, Fe) over volatile element Zn for our detection and clear subset. The USMgII absorbers galaxies exhibit gas depletion patterns broadly consistent with DLAs \citep{Mas-Ribas2017ApJ...846....4M}, and halo of the Milky Way reported by \citet{Welty1999ApJ...512..636W}. We find a similar depletion pattern along detection and clear-sightlines. For the clear sightlines, the \feii\ is depleted by a factor of two higher than the DLAs suggesting a substantial dust content, however, more data would be helpful to test it further. Given that the gas along clear sightlines might be associated with a faint galaxy at a small impact parameter, a halo of a massive galaxy at a large impact parameter, and/or intra-group gas, a follow-up study tracing the large-scale environment would help explore it further.

\subsection{Environment of USMgII absorbers}
\begin{figure}
    \centering
\includegraphics[width=0.48\textwidth]{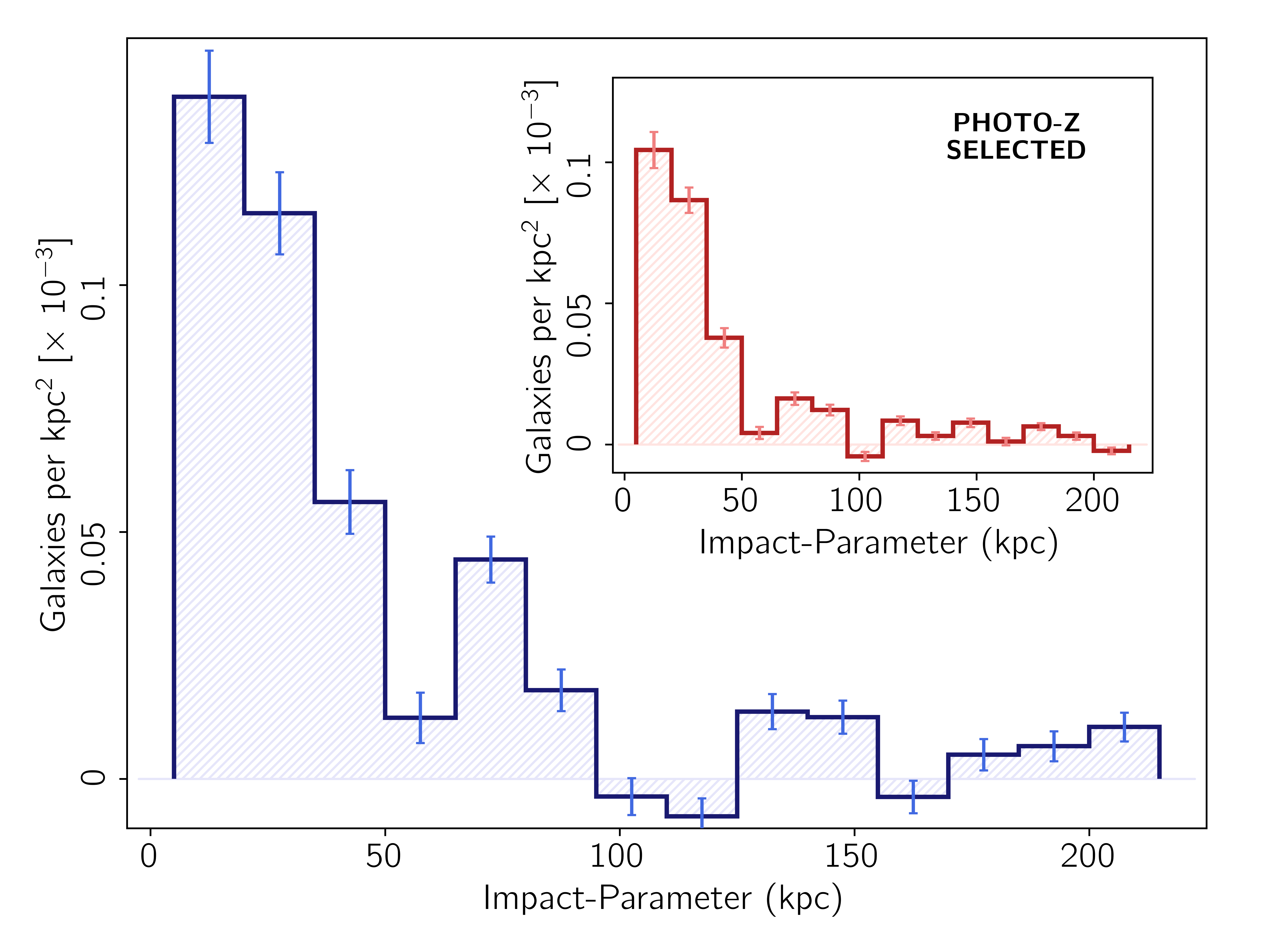}
    \caption{Excess surface density of galaxies along USMgII sightlines as a function of impact parameter with respect to a control quasar set of similar redshift and color. A clear overdensity of galaxies is evident for $\rho \lesssim 50$kpc. The inset shows the same for galaxies selected based on photometric redshift.} 
    \label{fig:QSOENV}
\end{figure}

Finally, we test if the USMgII absorber sightlines on average exhibit more sources at absorber redshift compared to the randomly selected field. For this, we select a control sample of ten quasars (with a minimum of 5 quasars), from the SDSS quasar catalog along each USMgII sightline that are matched in both redshift and magnitude, ensuring uniform levels of reddening and source density.  Using the photometric sources within 30 arcsec around quasar from deep HSC imaging, we estimate the excess surface density of galaxies out to 200 kpc, which is shown in Figure~\ref{fig:QSOENV}. A factor three higher overdensity of galaxies along USMgII sightlines is seen at $\rho \lesssim 50$~kpc in comparison to larger impact parameters. In the present study, the direct detection of \detectionpercentminfourfilter\% galaxies in proximity to quasars at an average impact parameter of within $\rho \le \ipmediankpc$ kpc supports the excess galaxy density. \citet{Nestor2007ApJ...658..185N} have reported an excess number of galaxies, with every sightline hosting at least one galaxy at impact parameter $\rho \lesssim 40$kpc and luminosity L$\gtrsim$0.3L$^{\star}$. \par

We further performed the above exercise by matching the absorber redshift with the HSC photometric redshift of galaxies within $\sigma [\Delta z/ (1+z)] \le$ 0.1 and found an excess at impact parameters of $\lesssim 50$kpc (see inset in  Fig~\ref{fig:QSOENV}). Note that, the excess overdensity is a lower limit since it is an average over a wide redshift range, resulting in varying mass limits. The MAGG survey \citep{Dutta2020MNRAS.499.5022D}, mapping the gas-galaxy association up to $\sim$ 200 kpc from quasars, revealed that $\sim$ 67\% of absorbers are associated with multiple galaxies. Taking advantage of HSC photometric redshifts, we explore the gas-galaxies association within an impact parameter of 200 kpc. The HSC photometric redshift-based association of galaxies with USMgII absorbers, including our detections, reveals that about \singlehostabsorberincourspercent\% of absorbers are likely produced by a single galaxy. Whereas, \doublehostabsorberincourpercent\% of the absorbers are associated with galaxy pair and \multihostabsorberincourspercent\% of the absorbers likely tracing the merger or group environments, which is consistent with the MAGG survey.   It suggests that apart from the strong outflows the wide velocity spread in USMgII absorbers originates from the galaxy-galaxy interaction \citep[see][]{Guha2022MNRAS.513.3836G}, and reiterates the importance of a large-scale IFU survey to study the gas-galaxy connection. The excess surface density of galaxies and a large fraction (\nondetectionpercentminfourfilter\%) of sightlines without a nearby potential host galaxy (i.e. clear cases) in the present study suggest that, in addition to the gas in-outflows, the galaxy pairs and intra-group medium may likely contribute to the kinematically complex absorption profile of USMgII systems.

\section{Conclusion}

Utilizing the deep, high-resolution optical imaging data provided by the Subaru HSC-SSP survey, we investigated the characteristics of galaxies associated with USMgII absorbers. Augmented by deep IR imaging data from the VISTA survey and spectral data from the SDSS survey, we have identified \detection\ USMgII galaxies in proximity to quasars across a broad redshift range of $\allabszmin\ \le z \le \allabszmax$. The morpho-kinematic analysis of these elusive systems resulted in the following key findings:

1. We detect the USMgII galaxies in proximity to the quasar, with an impact parameter ranging between $\ipstartkpc\ \le \rho \le \ipendkpc$\ kpc, in approximately \detectionpercent\% of cases. For the average impact parameter of $< \kpconepfivearcsecavgz$\ kpc, considering the availability of a minimum of 4 HSC passbands we obtain a detection rate of \detectionpercentminfourfilter\%. For \detectionoiidetected\ systems, we detect \oii\ nebular emission at $\ge 2 \sigma$ level, while a strong \oii\ emission is observed in the stacked spectrum for the entire subset, likely originated in the outflows or extended gaseous disk. However, for \nondetectionpercent\% (\nondetection\ out of \totalminimumfourfilteroroiidetected) of absorbers, no counterpart is detected in close quasar proximity. The quasar sightline hosting USMgII absorber, compared to a control quasar sample of similar redshift and color, shows a factor three higher galaxy surface density of $\sim$\excessgalaxiesperkpcsquare\ galaxies per kpc$^{2}$, especially at an impact parameter of $\lesssim 50$kpc.

2. The galaxies hosting USMgII absorbers in our sample exhibit a broad stellar mass distribution, ranging from $\rm \minstellarmassdetected\ \le log\ M_{\star} [M_{\odot}] \le  \maxstellarmassdetected$, with an average star formation rate of \meansfrdetected $~\rm M_{\odot}\ yr^{-1}$. These galaxies generally exhibit the star formation rates of a typical main sequence star-forming galaxies. About \starburstpercent\% of USMgII galaxies show a starburst nature whereas all of them meet the threshold star formation rate for launching strong outflows.

3. In contrast to the observed anti-correlation between \ew\ and  $\rho$ in the literature, the USMgII absorbers show a significant scatter in \ew\  out to $\sim$25 kpc. Considering the bi-conical wind geometry, the USMgII systems with large azimuthal angles, tracing the outflows, exhibit near-constant \ew\  up to large impact parameters. In addition, the wind subset shows preferentially stronger \ew\ than the ones produced in the extended disk. A strong preference of the USMgII systems along the direction of outflow ($\alpha \ge 50^{\circ}$) or extended galactic disk ($\alpha \le 30^{\circ}$) and the large scatter in \ew\ as a function of $\rho$ supports their wind and disk origin. 

Finally, the non-zero \oii\ emission along clear sightlines, yet no discernible stellar counterpart in the deep Subaru HSC images with a typical depth of $r_{mag} \sim 26$ are likely potential dark galaxies. This is further corroborated by a few instances where no galaxy counterpart is observed in any of the $g,\ r,\ i,\ z,\ y$ bands, in contrast, a faint galaxy counterpart is detected in coadded multi-band HSC image frames (Das et al., in preparation). The USMgII origin may also related to satellites of $\rm logM_{\star} < 10^8 M_{\odot}$, around central $\rm logM_{\star} < 10^9 M_{\odot}$ galaxy within a typical $\Delta v$ ranging from $\sim70 - 250$\kms, which account for $\sim$40 percent absorber fraction up to the virial radius in TNG 50 simulation \citep[see,][]{Weng2024MNRAS.527.3494W}. The multi-band imaging and spectral follow-up of these sightlines will contribute to understanding the effects of strong outflows on galaxy growth across a range of masses.

\begin{acknowledgements}
LCH was supported by the National Science Foundation of China (11991052, 12233001), the National Key R\&D Program of China (2022YFF0503401), and the China Manned Space Project (CMS-CSST-2021-A04, CMS-CSST-2021-A06). HMY was partially supported by the Research Fund for International Young Scientists of NSFC (11950410492) and JSPS KAKENHI Grant Number JP22K14072. 

The Hyper Suprime-Cam (HSC) collaboration includes the astronomical communities of Japan and Taiwan, and Princeton University. The HSC instrumentation and software were developed by the National Astronomical Observatory of Japan (NAOJ), the Kavli Institute for the Physics and Mathematics of the Universe (Kavli IPMU), the University of Tokyo, the High Energy Accelerator Research Organization (KEK), the Academia Sinica Institute for Astronomy and Astrophysics in Taiwan (ASIAA), and Princeton University. Funding was contributed by the FIRST program from Japanese Cabinet Office, the Ministry of Education, Culture, Sports, Science and Technology (MEXT), the Japan Society for the Promotion of Science (JSPS), Japan Science and Technology Agency (JST), the Toray Science Foundation, NAOJ, Kavli IPMU, KEK, ASIAA, and Princeton University. 

This paper makes use of software developed for Vera C. Rubin Observatory. We thank the Rubin Observatory for making their code available as free software at http://pipelines.lsst.io/.

This paper is based on data collected at the Subaru Telescope and retrieved from the HSC data archive system, which is operated by the Subaru Telescope and Astronomy Data Center (ADC) at NAOJ. Data analysis was in part carried out with the cooperation of Center for Computational Astrophysics (CfCA), NAOJ. We are honored and grateful for the opportunity of observing the Universe from Maunakea, which has the cultural, historical and natural significance in Hawaii. 

The Pan-STARRS1 Surveys (PS1) and the PS1 public science archive have been made possible through contributions by the Institute for Astronomy, the University of Hawaii, the Pan-STARRS Project Office, the Max Planck Society and its participating institutes, the Max Planck Institute for Astronomy, Heidelberg, and the Max Planck Institute for Extraterrestrial Physics, Garching, The Johns Hopkins University, Durham University, the University of Edinburgh, the Queen’s University Belfast, the Harvard-Smithsonian Center for Astrophysics, the Las Cumbres Observatory Global Telescope Network Incorporated, the National Central University of Taiwan, the Space Telescope Science Institute, the National Aeronautics and Space Administration under grant No. NNX08AR22G issued through the Planetary Science Division of the NASA Science Mission Directorate, the National Science Foundation grant No. AST-1238877, the University of Maryland, Eotvos Lorand University (ELTE), the Los Alamos National Laboratory, and the Gordon and Betty Moore Foundation.

Funding for the Sloan Digital Sky Survey IV has been provided
by the Alfred P. Sloan Foundation, the U.S. Department of
Energy Office of Science, and the Participating Institutions.
SDSS-IV acknowledges support and resources from the Center
for High-Performance Computing at the University of Utah. The
SDSS website is www.sdss.org.

SDSS-IV is managed by the Astrophysical Research Consortium for the Participating Institutions of the SDSS Collaboration, including the Brazilian Participation Group, the Carnegie
Institution for Science, Carnegie Mellon University, the Chilean
Participation Group, the French Participation Group, Harvard Smithsonian Center for Astrophysics, Instituto de Astrofísica de
Canarias, The Johns Hopkins University, Kavli Institute for the
Physics and Mathematics of the Universe (IPMU)/University of
Tokyo, the Korean Participation Group, Lawrence Berkeley
National Laboratory, Leibniz Institut für Astrophysik Potsdam
(AIP), Max-Planck-Institut für Astronomie (MPIA Heidelberg),
Max-Planck-Institut für Astrophysik (MPA Garching), MaxPlanck-Institut für Extraterrestrische Physik (MPE), National
Astronomical Observatories of China, New Mexico State
University, New York University, University of Notre Dame,
Observatário Nacional/MCTI, The Ohio State University,
Pennsylvania State University, Shanghai Astronomical Observatory, United Kingdom Participation Group, Universidad Nacional
Autónoma de México, University of Arizona, University of
Colorado Boulder, University of Oxford, University of Portsmouth, University of Utah, University of Virginia, University of
Washington, University of Wisconsin, Vanderbilt University, and
Yale University. 
\end{acknowledgements}

%


%
%

\bibliographystyle{aa} 
\bibliography{aa} 
\end{document}